\documentclass[10pt,journal,compsoc]{IEEEtran}

\usepackage{colortbl}
\usepackage{url}
\usepackage{enumitem}
\usepackage[nomessages]{fp}

\usepackage{enumerate}
\usepackage{footnote}
\makesavenoteenv{tabular}
\makesavenoteenv{table} 

\usepackage{tikz} 
\usetikzlibrary{decorations.text}
\usetikzlibrary{patterns, automata}
\usetikzlibrary{calc,trees,positioning,arrows,chains,shapes.geometric,%
decorations.pathreplacing,decorations.pathmorphing,shapes,%
matrix,shapes.symbols}
\usetikzlibrary{arrows}

\usepackage{subcaption}

\usepackage{multirow}    
\usepackage{hhline}      
\usepackage{algorithm} 
\usepackage{algpseudocode}
\usepackage{enumerate}
\usepackage{graphics}
\usepackage[hidelinks]{hyperref}
\usepackage{listings}
\usepackage{soul} 
\usepackage{color}
\definecolor{KWColor}{rgb}{0.37,0.08,0.25}
\definecolor{CommentColor}{rgb}{0.133,0.545,0.133}
\definecolor{StringColor}{rgb}{0,0.126,0.941}
\usepackage{pgfplots}
\pgfplotsset{compat=1.4}

\usepackage{framed}
\usepackage{mdframed}

\usepackage{graphicx}

\usepackage{flushend}

\usepackage{pifont}
\newcommand{\cmark}{\ding{51}}%
\newcommand{\xmark}{\ding{55}}%

\usepackage{amssymb,amsmath}
\usepackage{ifthen}
\usepackage{color}

\usepackage[most]{tcolorbox}

\newboolean{showcomments}
\setboolean{showcomments}{true}
\ifthenelse{\boolean{showcomments}}
{ 
  \newcommand{\mynote}[2]{
  \fbox{\bfseries\sffamily\scriptsize#1}
  {\small$\blacktriangleright$\textsf{\emph{#2}}$\blacktriangleleft$}}
}{ 
  \newcommand{\mynote}[2]{}
}

\usepackage{rotating}

\newmdenv[
topline=false,
bottomline=false,
rightline=false,
skipabove=0pt,
skipbelow=0pt,
linewidth=1pt,
]{definition}

\newmdenv[
linewidth=1pt,
]{textfig}

\newcommand{\repo}[1]{{RePack}}

\newcommand{\numofpapers}[1]{{57}}
\definecolor{lightgray}{gray}{0.85}

\begin{document}

\title{Rebooting Research on Detecting Repackaged Android Apps: Literature Review and Benchmark}

\author{Li~Li,~
        Tegawend\'e~F.~Bissyand\'e,~
        Jacques~Klein
\IEEEcompsocitemizethanks{
\IEEEcompsocthanksitem
L. Li is with the Faculty of Information Technology, Monash University, Australia.
\IEEEcompsocthanksitem
T. Bissyand\'e and J. Klein are with the Interdisciplinary Centre for Security, Reliability and Trust, University of Luxembourg, Luxembourg.}
\thanks{Manuscript received XXX; revised XXX.}}

\IEEEtitleabstractindextext{%

\begin{abstract}

Repackaging is a serious threat to the Android ecosystem as it 
deprives app developers of their benefits, 
contributes to spreading malware on users' devices, 
and increases the workload of market maintainers.
In the space of six years, the research around this specific issue has produced \numofpapers{} approaches which
do not readily scale to millions of apps or are only evaluated on private datasets without, in general, tool support available to the community.
Through a systematic literature review of the subject,
we argue that the research is slowing down, where many state-of-the-art approaches have reported
high-performance rates on closed datasets, which are unfortunately difficult to replicate and to compare against.
In this work, we propose to reboot the research in repackaged app detection by providing a literature review that summarises the challenges and current solutions for detecting repackaged apps and 
by providing a large dataset that supports replications of existing solutions and implications of new research directions.
We hope that these contributions will re-activate the direction of detecting repackaged apps and spark innovative approaches going beyond the current state-of-the-art.

\end{abstract}


}

\maketitle

\section{Introduction}
\label{sec:introduction}

Mobile applications, especially Android apps, are straightforward to reverse engineer, 
copy and resubmit to markets~\cite{Gibler:2013:AEL:2462456.2464461}. 
The Android application packaging system indeed relies on the ZIP open compression format to archive apps' resource files and 
decompilable bytecode, making it easy for anyone with the adequate tool support 
to unpack any app, modify its contents, and repackage it. {\em Repackaging} is thus
a common threat to the Android ecosystem, where it is used by {\em plagiarists} 
who {\em clone} apps from other developers, e.g., in order to redirect advertisement 
revenue~\cite{Gibler:2013:AEL:2462456.2464461,crussell2012attack}, and by malware writers who {\em piggyback} malicious payloads on popular 
apps to spread malware~\cite{li2016ungrafting}.

Android app repackaging has been raised as a serious problem by various authors in the literature as well as by different stakeholders in the app development industry. 
For example, large-scale application plagiarism~\cite{Ankeny12} has led to the shutdown of several non-official app markets. Last year, Ustwo Games, developer of the popular ``Monument Valley'' game, has reported that only 5\% of Monument Valley installations on Android are paid for~\cite{ustwo_games}, with various copies being available from different ``authors'' in the same market. 
 Very recently, the famous Pokemon Go app has been repackaged in different ways and for different reasons as mentioned in the Lookout blog~\cite{pokemon_go}.  
 In a different category, Jung et al.~\cite{Jung2013} have presented a serious case of repackaging attacks 
 on Korea's Banking Android apps, demonstrating how it was possible to redirect money 
 transfers without having to illegally obtain any of the sender's personal information 
 such as bank accounts.
 
 To limit repackaging and its impacts, different steps must be taken by all concerned parties.
 For example,  developers may explore different techniques for watermarking their apps and denying some functionalities when it becomes obvious that the running copy is a cloned version~\cite{Zhou:2013:AWA:2484313.2484315,ren2014droidmarking}. Nevertheless, 
 most of the workload is carried by market maintainers who must employ powerful tools to catch repackaged apps in a fast and accurate way so as to remove them from markets~\cite{wang2018android}. 
 A number of research studies in the literature have investigated a variety of repackaging 
 detection approaches without convincingly demonstrating that the problem is now well addressed.

In this paper, we revisit the state of research on repackaged app detection, insisting on 
the practical challenges that the community must focus on for implementing effective
repackaged app detection solutions for Android markets. Overall we make the following 
contributions: 

\begin{itemize}
\item We propose a systematic review of the state-of-the-art literature on repackaged app detection and highlight their shortcomings in terms of the impracticality of the approaches, lack of reproducibility, and suboptimal evaluation scenarios.
\item We build the \repo{} dataset and release it to the community to encourage proper assessment of repackaged app detection approaches.
Our work builds upon the popular AndroZoo repository~\cite{allix2016androzoo,li2017androzoo++}, which can serve as an exchange repository for describing one's dataset using the hash values of apps.
We also enumerate research directions that the community should take up for advancing the state-of-the-art in the topic.
\end{itemize}

The remainder of this paper is organised as follows. Section~\ref{sec:state} clarifies 
the terminology used, and explains the procedure for the Systematic
 Literature Review (SLR) that we have conducted on the topic of Android  repackaged app detection.
 Section~\ref{sec:challenges} discusses the prominent challenges in taming  
 app repackaging. Then, Section~\ref{sec:solutions} summarises the different 
 contributions made in the literature and  highlight some issues in their approaches and evaluations.
 Section~\ref{sec:future_directions} describes our efforts in addressing some important 
 challenges. Notably, we discuss the construction of a large dataset of repackaging 
 pairs from the  AndroZoo app repository. 
 Sections~\ref{sec:priority} and~\ref{sec:validity} discuss priority research directions and potential threats to validity of this study respectively.
 Section~\ref{sec:related_work} enumerates closely related work and 
 Section~\ref{sec:conclusion} finally provides concluding remarks.
 			
All artefacts of our research are available in the \repo{} repository at:
\begin{center}
	{\tt \url{https://github.com/serval-snt-uni-lu/RePack}}
\end{center}

\section{Literature Search}
\label{sec:state}
In this section, we provide introductory information on the lightweight Systematic Literature Review (SLR) that we have 
performed to assess the advances that were made in the area of repackaged app detection.
We first clarify the terminology used in the field before giving statistics on the collected literature corpus.

\subsection{Terminology}
Several terms are used in the literature, notably in paper titles and abstracts, to indicate actions
that somehow involve a repackaging process:

{\em Repackaging} refers to the core process of unpacking a software package, then repackaging it after
a probable modification of the decompiled code and/or of other resource files (e.g. logos, Permission list, etc.). Because
all Android app packages (APKs) are signed with the developer certificate, a repackaging pair,
formed by an original app and its repackaged version, can be differentiated by their checksums even 
when no modification of the code has been performed by the repackager, who should be different from the original app developer. 
Following this principle, \textbf{we consider in this work two apps as a repackaged app pair as long as (1) they share at least 80\% of the code (i.e., code similarity exceeds 80\%) and (2) they are signed by different developers}.

{\em Cloning} is the process of building a software by reverse engineering another software or by reimplementing
it based on documentation or usage experience. 
Theoretically, cloning is different from repackaging because it does not need to package an app based on its cloning version while repackaging always form the app based on its original counterpart.
Nevertheless, in the Android ecosystem, this difference is negligible as it is straightforward to clone an app via repackaging and, in most cases, the whole app code (rather than partial code) is manipulated.

{\em Reusing} is the action of producing apps from existing code rather than developing from scratch.
The existing code can vary from small parts like several methods to big parts such as whole app functionalities. 

{\em Plagiarism} consists in wrongfully appropriating the work of another developer, e.g., by cloning her/his APK to
benefit from, for example, advertisement revenues. 
Comparing to the term reusing, plagiarism emphases on the part that the cloned code is wrongfully leveraged.

{\em Piggybacking} is defined in the literature as a malware development activity where a given benign app is repackaged to include a malicious payload. Piggybacked apps thus constitute a subset of repackaged apps.

{\em Camouflage} is a technique used by malware writers to trick users into installing malware sample, which is
presented as a well-known popular app, e.g., by repackaging an app to replace its main functionality with the malicious
implementation.
In this work, we consider camouflage as a special case of piggybacking, where the app interfaces are not modified (to keep the same looks) but the app code has been manipulated.

\subsection{Systematic Literature Review (SLR) Methodology}
Our work evolves around an investigation of the state-of-the-art research on repackaged app detection.
We search for the relevant literature in a systematic way, following the guidelines provided by Keele~\cite{keele2007guidelines}, 
and Brereton et al.~\cite{brereton2007lessons}.
Thus, in a first step, after outlining the relevant research questions, we search for potential related work (up to the end of 2017) in four well-known online repositories: ACM Digital Library, IEEE Xplore, SpringerLink, and ScienceDirect. We use two groups of keywords (in the form of regular expression) enumerated in Table~\ref{tab:search_keywords}.
The search string\footnote{(android OR mobile OR *phone*) AND (clon* OR repackag* OR piggyback* OR plagiari* OR reus*)} is formed as a conjunction {\tt g$_1$ AND g${_2}$} where {\tt g$_1$} and {\tt g$_2$} are themselves 
formed each as a disjunction of the keywords respectively of groups G1 and G2. 
The goal of this step is to collect as many related papers as possible, taking into account most well-recorded proceedings. We consolidate\footnote{Online repository search engines often list irrelevant results presence potential irrelevant articles~\cite{li2016static,kong2018automated}.} the collected list of relevant work by manually going through all the papers, examining the title and abstract, to ensure that they deal with repackaged app detection. 
Following the same guidelines suggested by Barbara Kitchenham~\cite{kitchenham2004procedures}, short papers\footnote{Less than five pages in double column or nine pages in single column.} such as the one presented by Ayush Kohli~\cite{kohli2017decisiondroid} will not be considered in this study.

In a second step, we perform a backwards-snowballing on the remaining papers in an attempt to account for influential papers that may not have been recorded in the aforementioned repositories or that did not mention the used keywords. To that end, we carefully read the
related work section of the papers collected at the end of the first step.

\begin{table}[!h]
\centering
\caption{Repository Search Keywords.}
\label{tab:search_keywords}
\resizebox{\linewidth}{!}{
\begin{tabular}{r l}
\hline
{\bf Group} ({\tt AND}) & {\bf Keywords} ({\tt OR}) \\
\hline
G1		&	android, mobile, *phone* \\
G2		& 	clon*, repackag*, piggyback*, plagiari*, reus*, camouflag* \\
\hline
\end{tabular}
}
\end{table}

At the end of the SLR search, we had collected \numofpapers{} papers that present work dealing, in one way or another,
and to any extent, with research on Android app repackaging.
Table~\ref{tab:full_list_papers} enumerates all the papers, 
highlighting their publication year, publication venues and the
accompanying tool name.

\begin{table}[!h]
\centering
\caption{Full List of Collected and Examined Papers}
\label{tab:full_list_papers}
\resizebox{\linewidth}{!}{
\begin{tabular} { ll r }
\hline
Tool/Reference & Year & Venue$^\alpha$\\ 
\hline

CodeMatch~\cite{glanz2017codematch} & 2017 & ESEC/FSE \\
\rowcolor{lightgray} DR-Droid2~\cite{tian2017detection} & 2017 & TDSC (J) \\
DAPASA~\cite{fan2017dapasa} & 2017 & TIFS (J) \\
\rowcolor{lightgray} FUIDroid~\cite{lyu2017efficient} & 2017 & MISY (J) \\
APPraiser~\cite{ishii2017appraiser} & 2017 & IEICE TIS (J)\\
\rowcolor{lightgray} RepDroid~\cite{yue2017repdroid} & 2017 & ICPC\\
SimiDroid~\cite{li2017simidroid} & 2017 & TrustCom\\
\rowcolor{lightgray} GroupDroid~\cite{marastoni2017groupdroid} & 2017 & SSPREW@ACSAC (W)\\

CLANdroid~\cite{linares2016onautomatically}&2016&ICPC\\
\rowcolor{lightgray} DR-Droid~\cite{tian2016analysis}&2016&MoST@S\&P (W)\\ 
DroidClone~\cite{alam2016droidclone}&2016&DICTAP\\
\rowcolor{lightgray} FSquaDRA2~\cite{gadyatskaya2016evaluation} & 2016    &	NordSec \\
Li et al.~\cite{li2016investigation}&2016&SANER\\ 
\rowcolor{lightgray} Niu et al.~\cite{niu2016clone} & 2016 & ICSAI\\ 
RepDetector~\cite{guan2016semantics}&2016&ESSoS\\ 
\rowcolor{lightgray} SUIDroid~\cite{lyu2016suidroid}    &    2016    &    TrustCom\\

Kim et al.~\cite{kim2015detecting} & 2015 & ASE (J)\\
\rowcolor{lightgray} AndroidSOO~\cite{gonzalez2015exploring} & 2015 &  EuroSec@EuroSys (W) \\
AndroSimilar2~\cite{faruki2015androsimilar} & 2015 & JISA\\
\rowcolor{lightgray}Chen et al.~\cite{chen2015detecting}&2015&JCST (J)\\ 
 DroidEagle~\cite{sun2015droideagle}&2015&WiSec\\ 
\rowcolor{lightgray} ImageStruct~\cite{jiao2015rapid}&2015&ISPEC\\ 
 MassVet~\cite{chen2015finding}&2015&USENIX Security\\ 
\rowcolor{lightgray} PICARD~\cite{aldini2015detection}&2015&CCPE (J)\\ 
 Soh et al.~\cite{soh2015detecting}&2015&ICPC\\ 
\rowcolor{lightgray} Wu et al.~\cite{wu2015detect}&2015&SCN (J)\\ 
WuKong~\cite{wang2015wukong}&2015&ISSTA\\ 

\rowcolor{lightgray}AnDarwin2~\cite{crussell2014andarwin}&2014&TMC (J)\\
AndRadar~\cite{lindorfer2014andradar}&2014&DIMVA\\ 
\rowcolor{lightgray}Chen et al.~\cite{chen2014achieving}&2014&ICSE\\ 
 DIVILAR~\cite{zhou2014divilar}&2014&CODASPY\\ 
\rowcolor{lightgray} DroidKin~\cite{gonzalez2014droidkin}	& 2014  & SecureComm \\
 DroidLegacy~\cite{deshotels2014droidlegacy}&2014&PPREW@POPL (W)\\ 
\rowcolor{lightgray} DroidMarking~\cite{ren2014droidmarking}&2014&ASE\\ 
 DroidSim~\cite{sun2014detecting}&2014&IFIP SEC\\ 
\rowcolor{lightgray} FSquaDRA~\cite{zhauniarovich2014fsquadra}&2014&DBSec\\ 
 Kywe et al.~\cite{kywe2014detecting}&2014&ICISC\\ 
\rowcolor{lightgray} Linares-Vásquez et al.~\cite{linares2014revisiting}&2014&MSR\\  
 PLayDrone~\cite{viennot2014measurement}&2014&SIGMETRICS\\ 
\rowcolor{lightgray} ResDroid~\cite{shao2014towards}&2014&ACSAC\\ 
Ruiz et al.~\cite{ruiz2014large}&2014&IEEE Software (J)\\ 
\rowcolor{lightgray} ViewDroid~\cite{zhang2014viewdroid}&2014&WiSec\\

AdRob~\cite{gibler2013adrob}&2013& MobiSys\\ 
\rowcolor{lightgray} AnDarwin~\cite{crussell2013andarwin}&2013 &ESORICS\\ 
 AndroSimilar~\cite{faruki2013androsimilar} & 2013 & SIN\\
\rowcolor{lightgray} AppInk~\cite{zhou2013appink}&2013& AsiaCCS\\ 
AppIntegrity ~\cite{vidas2013sweetening}&2013&CODASPY\\
\rowcolor{lightgray} DroidAnalytics ~\cite{zheng2013droidanalytics}&2013 &TrustCom\\ 
PiggyApp~\cite{zhou2013fast}&2013&CODASPY\\  
\rowcolor{lightgray}  SCSdroid~\cite{lin2013identifying}&2013&CompSec (J)\\ 

Androguard~\cite{desnos2012android}&2012& HICSS\\  
\rowcolor{lightgray} DNADroid~\cite{crussell2012attack}&2012& ESORICS\\  
DroidMat~\cite{wu2012droidmat}&2012& AsiaJCIS\\
\rowcolor{lightgray} DroidMOSS~\cite{zhou2012detecting}&2012 &CODASPY\\ 
JuxtApp~\cite{hanna2012juxtapp}&2012& DIMVA\\ 
\rowcolor{lightgray} Potharaju et al.~\cite{potharaju2012plagiarizing}&2012& ESSoS\\ 
Ruiz et al.~\cite{ruiz2012understanding}&2012 &ICPC\\

\hline
\multicolumn{3}{l}{$^\alpha$: (J) and (W) stand for Journal and Workshop venues respectively}
\end{tabular}
}
\end{table}

\subsection{Statistics on state-of-the-art work}
\begin{figure}[!h]
    \centering 
    	\includegraphics[width=\linewidth]{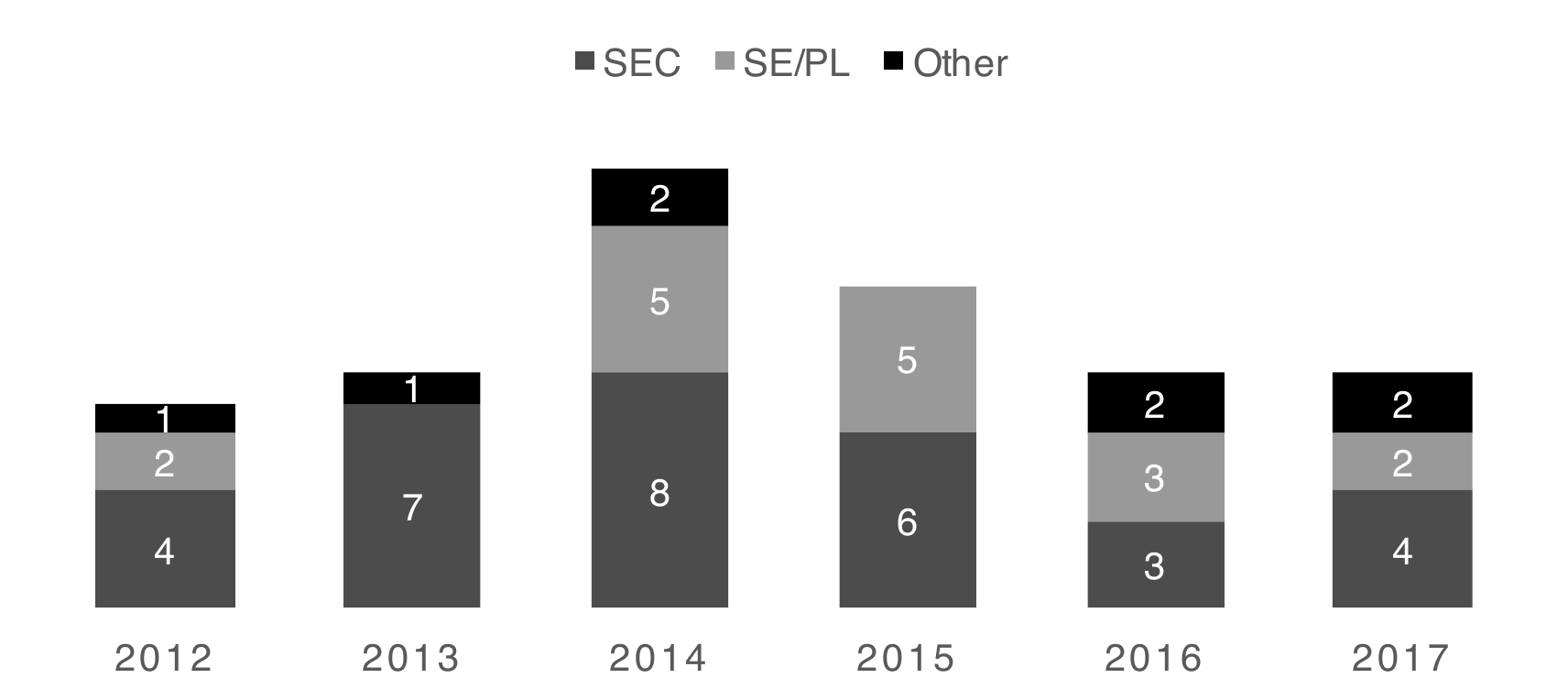}
    \caption{Distributions of the State-of-the-art literature per Year.}
    \label{fig:trend1}
\end{figure}

The research topic around repackaged apps has been initiated in the Android community after
a presentation of Desnos and Gueguen~\cite{desnos2011android} at Black Hat, Abu Dhabi 2011, where they discussed
Android app reverse engineering and decompilation, and the associated security implications.
Fig.~\ref{fig:trend1}  illustrates some statistical trends of the research publications 
on Android repackaged app detection.

It appears from the collected data that repackaged app detection has been tackled first and mostly by security researchers. 
Then, Software Engineering researchers have picked up on the problem, leading to a peak of
publications in 2014. After 2014, the volume of published research started to steadily decrease, 
although no new data has shown that the problem has been solved in practice.

\FPeval{\percent}{clip(round(100*11/\numofpapers{}, 0))}

It is noteworthy that only $\percent\%$  (11 out of \numofpapers{}) of the state-of-the-art work
have been archived in a Journal volume, suggesting that very few extensive investigations
into the problem are available. Conference proceedings, which provide a faster visibility
of researcher's work on a competitive topic such as Android, account for over $80\%$ of the publications.
The four workshop papers~\cite{deshotels2014droidlegacy,gonzalez2015exploring,tian2016analysis, marastoni2017groupdroid} that we have identified in the SLR
are providing radically new approaches to the problem of repackaged
apps detection, but only focus on repackaged malware.

Except for the considered approaches that explicitly target the detection of repackaged Android apps, our literature search also identified several papers focusing on detecting third-party libraries~\cite{li2017libd, ma2016libradar, backes2016reliable}.
Although these papers are not considered in this work, we believe their approaches can generally be adapted to detect repackaged apps as well.
For example, Li et al.~\cite{li2017libd}  have introduced LibD to identify third-party libraries, including multi-package ones, which are categorised based on the internal code dependencies of candidate libraries.
Interested readers are encouraged to follow their research paper for more details.

\section{Overview of Challenges in Repackaged App Detection}
\label{sec:challenges}
Before detailing the different solutions presented in the literature, we propose 
to review the challenges that researchers should seek to address in repackaged app detection.
These challenges are brought up by the realities in the app industry, practical requirements
for assessing a detection approach, as well as specificities of Android development.

\textbf{(1)Meeting market-scale constraints.}\\
Android developers have produced millions of apps distributed in several markets, raising
scalability issues in the detection of repackaged apps. 
In this work, we consider that the scalability of detecting repackaged apps is referred to two challenges: (1) combinatorial explosion and (2) impractical aspect, e.g., to have the original app counterparts in the dataset.

Regarding combinatorial explosion, let us take Google Play as an example, the official Android app market has hosted already over 3 million Android apps\footnote{\url{https://www.appbrain.com/stats/number-of-android-apps}}, while several
alternative markets such as AppChina have also passed the one-million mark. A simple and intuitive,
pairwise similarity comparison between apps (combinatorial explosion) is thus not scalable in practice. For example,
using such an approach to detect repackaged apps in Google Play alone, one would need to
perform about $C^{2}_{3*10^6}$ comparisons. If we consider a computing platform with 
10 cores each starting 10 threads to compare pairs of apps in parallel, 
it would still take several months to complete the analysis when optimistically assuming 
that each comparison would take about 1ms.

From a practical point of view, unfortunately, the problem is further exacerbated by the fact that repackaged apps and their original counterparts are often hosted on different markets. 
A pairwise comparison approach would then require collecting as many apps as possible across all markets and repositories.
Failing to collect such app set would result in an unfair evaluation for repackaged app detection approaches.
Indeed, on the one hand, because of missing original apps, some repackaging detection approaches (e.g., pairwise-based approaches) would be unsuccessful for flagging some repackaged apps, although those approaches by themselves are capable of achieving that.
On the other hand, given a flagged repackaged app (e.g., by ML-based classifiers), not being able to identify its original counterpart in the testing app set does not necessarily mean the flagged app is a false alarm, because it could be the case that the testing app set is just not big enough, where the original counterpart happens to not be in it.  
 
\textbf{(2)Having a reference dataset.}\\
Despite the awareness in academia and industry on the problem of app repackaging, the
community lacks relevant datasets to support research work. Building a large and consistent ground truth
of repackaging pairs indeed requires substantial efforts. Unfortunately, unless such efforts are
made, we can hardly expect significant advances towards producing reliable approaches and tools for
addressing repackaged app detection. Indeed, on the one hand, in the absence of a reference dataset, which can serve as a pseudo ground truth, 
state-of-the-art work cannot be benchmarked and compared one against the other. On the other hand,
existing approaches that claim to be successful cannot convince the reader on the precision of their
techniques, since confirmation is manually performed by the authors and the detected repackaged apps are not disclosed to the community for additional checking.

\textbf{(3)Recognising the original app in a repackaging pair.}\\
Given a repackaging pair, constituted by two similar apps, one being a repackaged version of the other, it is commonly accepted
that it remains challenging to distinguish which app is the original~\cite{gibler2013adrob,zhou2013fast}. 
Instead, the literature often relies on heuristics such as app packaging/compilation time (e.g., timestamp of {\em classes.dex} ).
Although such heuristics are intuitive, they are not fully reliable in a sophisticated malware development scenario.
 Indeed, it is possible for malware writers to manipulate compilation time of their repackaged apps~\cite{li2018moonlightbox}. Yet,
 a number of state-of-the-art approaches depend on such heuristics to flag repackaged apps in the wild.

\textbf{(4)Accounting for potential obfuscation.}\\
Obfuscation is known to be effective to help developers to hide their code logic for preventing potential plagiarism.
Many obfuscation algorithms have been implemented in frameworks such as DexGuard~\cite{dexguard} and SandMark~\cite{sandmark} which are already used in the Android community. At the same time,
however, obfuscation can be leveraged by pirates and malware writers to evade the detection of their repackaging operations.
As shown by Huang et al.~\cite{huang2013framework}, most static approaches for repackaged app detection  are ineffective in the presence
of obfuscated apps.

\textbf{(5)Dealing with noise of common libraries.}\\
Common libraries, such as the popular {\em com.google.ads} and {\em com.revmob} advertisement libraries, which are extensively used across many apps, can significantly impact the effectiveness in repackaged app detection~\cite{li2016investigation}. Indeed, when common libraries are substantially larger
than core functionality code, a pairwise comparison approach can lead to false positives, presenting two different apps, but with similar libraries, as a repackaging pair. Similarly, when a large library is replaced during repackaging by another library, a
pairwise comparison will fail to detect the repackaging scenario, leading to a false negative.
To overcome this challenge, researchers build whitelists of common libraries that are filtered out during repackaged detection processes~\cite{chen2014achieving}. Unfortunately, it is also challenging to build an exhaustive list of common libraries.

\textbf{(6)Constructing and exploiting a call/dependency graph.}\\
Call and dependency graphs are appealing for repackaged app detection as they can abstract the behaviour implemented in a software
to allow effectively identifying similar behaviour~\cite{Liu06gplag:detection,Roy07asurvey}. 
Unfortunately, there are several challenges in constructing an 
Android call/dependency graph that will be reliable for comparison experiments. First, Android is event-based and most behavioural actions are performed via user-triggered events (e.g., clicking a button) or system events 
(e.g., incoming phone call), through callback methods. A call graph may not properly account for the sequences in which callback methods are called. Second, the Inter-component communication mechanism further involves the use
of callback methods to allow interaction among different parts of an app. Since those parts are not directly linked at the code level,
the constructed call/dependency graph of the app will eventually be incomplete, depending on the choice of starting point for
exploring the app. Finally, heavy use of reflection further complicates the building of sound call/dependency graphs~\cite{li2016droidra}.

The size of the graphs can also challenge the detection of repackaged apps. Indeed, when the graphs are small (e.g., with less than four nodes), comparisons often lead to numerous false positives~\cite{chen2014achieving}. When the graphs are very large, the challenge
of finding isomorphisms may become prohibitive.

\textbf{(7)Dealing with corner-cases.}\\
Besides the aforementioned challenges, some corner cases are often eluded in the literature of app analysis.
First, some apps cannot be decompiled  by popular Baksmali and Soot de-compilers.
Indeed, malware writers may intentionally include code which is specifically engineered to prevent such compilers to work~\cite{Yang2015Raid}.
Second, all apps are not strictly packaged with common assets: for example, 
some apps do not have layouts, which may cause failures for approaches that assume that apps always do.
Finally, app hardening techniques, where the main app code in \emph{classes.dex} is encrypted and loaded at runtime through 
Java Native Interface (JNI), also raises the bar for the research in repackaged app detection~\cite{zhang2015dexhunter}.

\textbf{(8)Dealing with legal issues.}\\
Aside from technical barriers, legal concerns, including copyright/licensing issue in exposing third-party apps and liability in redistributing malware can impact the advances in research on repackaged app detection. Indeed,
as previously warned by Bodden et al.~\cite{bodden2013schutzmassnahmen}, data security and user privacy on one side and intellectual property rights on the other side, are slowly emerging in the field of informational self-protection.
Thus, for example, researchers are often forced to hold back their dataset, hindering adequate comparisons that could lead to tangible improvement of the state-of-the-art. As suggested by Rasthofer et al.~\cite{rasthofer2015droidsearch}, there is a need in our community to analyze legal issues.

\begin{table*}[!t]
\centering
\caption{Summary of Examined Approaches.}
\label{tab:approaches}
\resizebox{1.0\linewidth}{!}{
\begin{tabular} { l l p{0.38\linewidth} c  c | c c c c}
\hline
Tool & Category & Features &  Dynamic	& Bytecode	& MS$^\alpha$ & GT$^\beta$ & OR$^\gamma$ & LN$^\delta$\\ 
\hline
AndroidSOO~\cite{gonzalez2015exploring} & Symptom discovery	& string offset order		&	&  	 & \cmark & &  & \\

\rowcolor{lightgray} CodeMatch~\cite{glanz2017codematch} & Similarity Comparison & code fuzzy hash & & \cmark & & \cmark & \cmark & \cmark \\
FUIDroid~\cite{lyu2017efficient} & Similarity Comparison & layout tree & & & & \cmark & \cmark & \cmark \\
\rowcolor{lightgray} APPraiser~\cite{ishii2017appraiser} & Similarity Comparison & resource files & & & & & \cmark & \\
RepDroid~\cite{yue2017repdroid} & Similarity Comparison & layout group graph & \cmark & & & \cmark & \cmark & \\
\rowcolor{lightgray} SimiDroid~\cite{li2017simidroid} & Similarity Comparison & method statements, resource files, components & & \cmark & & \cmark & & \\
GroupDroid~\cite{marastoni2017groupdroid} & Similarity Comparison & control flow graph & & \cmark & & \cmark & & \\

\rowcolor{lightgray} CLANdroid~\cite{linares2016onautomatically} & Similarity Comparison 	& Identifiers, APIs, Intents, Permissions, and Sensors 		&		& \cmark   	&  & \cmark & & \cmark \\
Li et al.~\cite{li2016investigation} 	&	Similarity Comparison 	& method-level signature 		&		& \cmark   	&  & & & \cmark \\
\rowcolor{lightgray} RepDetector~\cite{guan2016semantics}	&	Similarity Comparison	& inputs/outputs of methods			& 	& \cmark  &&& \cmark & \cmark \\
Wu et al.~\cite{wu2015detect}		&	Similarity Comparison	& HTTP distance						&  \cmark &	 \cmark  &&& \cmark & \cmark \\

\rowcolor{lightgray} FSquaDRA2~\cite{gadyatskaya2016evaluation}         & Similarity Comparison	& signature of resource files		& 	& 		 && \cmark & \cmark & \\
SUIDroid~\cite{lyu2016suidroid}    &    Similarity Comparison    & layout tree   &   &	   &&& \cmark  & \cmark \\
\rowcolor{lightgray} DroidClone~\cite{alam2016droidclone} &    Similarity Comparison    & control flow pattern    &   &	 \cmark  &&& \cmark  & \\
Niu et al.~\cite{niu2016clone}     &    Similarity Comparison    &  method-level signature   &   &	 \cmark  &&& \cmark & \\
\rowcolor{lightgray} AndroSimilar2~\cite{faruki2015androsimilar} &    Similarity Comparison    &  entropy of byte block   &   &	 \cmark  &&& \cmark & \\
AndroSimilar~\cite{faruki2013androsimilar} &    Similarity Comparison    &  entropy of byte block  &   &	 \cmark  &&& \cmark & \\

\rowcolor{lightgray} DroidEagle~\cite{sun2015droideagle}	&	Similarity Comparison	& visual resources 					&   & 	 &&\cmark & \cmark  &  \\
ImageStruct~\cite{jiao2015rapid}		&	Similarity Comparison	& images								&  	&   &&&  &  \\
\rowcolor{lightgray} Soh et al.~\cite{soh2015detecting}	&	Similarity Comparison	& user interfaces					&  \cmark		&	 \cmark  && \cmark & \cmark & \cmark \\
Chen et al.~\cite{chen2015detecting}	&	Similarity Comparison	& method-level signature powered by NiCad~\cite{cordy2011nicad}	&  			& \cmark 	 &&&& \\
\rowcolor{lightgray} MassVet~\cite{chen2015finding}		&	Similarity Comparison			& centroid of UI structures, method-call graphs	& 	& \cmark 	 && \cmark & \cmark & \cmark \\	
DroidKin~\cite{gonzalez2014droidkin}						&	Similarity Comparison	& meta-info and n-gram bytecode/opcode	&	& \cmark 	 &&& \cmark & \\
\rowcolor{lightgray} Ruiz et al.~\cite{ruiz2014large}	&	Similarity Comparison	&	count-, set-, sequence-, and relationship-based objects &  &  \cmark  &&& \cmark & \cmark \\
Linares-V{\'a}squez et al.~\cite{linares2014revisiting}	& Similarity Comparison	&	count-, set-, sequence-, and relationship-based objects &  & \cmark   &&&& \\
\rowcolor{lightgray} Chen et al.~\cite{chen2014achieving}	&	Similarity Comparison	& centroid of control flow graph (CFG)	& 				&	 \cmark  && \cmark && \cmark \\
PLayDrone~\cite{viennot2014measurement}& Similarity Comparison	& signature of resource files		& 	& 		 &&&  &  \\
\rowcolor{lightgray} FSquaDRA~\cite{zhauniarovich2014fsquadra}& Similarity Comparison	& signature of resource files		& 	& 		 && \cmark &  &  \\
ViewDroid~\cite{zhang2014viewdroid}	&	Similarity Comparison	& ICC-based view graph				& 	&	 \cmark 		 &&& \cmark & \cmark \\
\rowcolor{lightgray} DroidSim~\cite{sun2014detecting}		&	Similarity Comparison	& component-based control-flow graph	& 		&	 \cmark 	 &&& \cmark & \cmark \\
AndRadar~\cite{lindorfer2014andradar}&	Similarity Comparison	& method-level signature					& 		&	 \cmark 	 &&&& \\
\rowcolor{lightgray} Kywe et al.~\cite{kywe2014detecting}	&	Similarity Comparison	& app name, description, icon, screenshot &   &   &&&  &  \\
PiggyApp~\cite{zhou2013fast}			&	Similarity Comparison	& APIs, permissions, Intents			& &	 \cmark  &&& \cmark & \cmark \\
\rowcolor{lightgray} DroidAnalytics~\cite{zheng2013droidanalytics} & Similarity Comparison 	& API sequences			&  &  \cmark  &&& \cmark & \cmark \\
AdRob~\cite{gibler2013adrob}			&	Similarity Comparison	& data-dependency graph	&  &  \cmark  &&&& \cmark \\
\rowcolor{lightgray} DroidMOSS~\cite{zhou2012detecting}	&	Similarity Comparison	& opcode sequences, 	developer certificate						& 		&	 \cmark 	 &&& \cmark & \cmark \\
JuxtApp~\cite{hanna2012juxtapp}		&	Similarity Comparison	& k-grams of opcode sequences		& 		&	 \cmark 	 &&&& \cmark \\
\rowcolor{lightgray} DNADroid~\cite{crussell2012attack}	&	Similarity Comparison	& program/data dependency graph		& 		&	 \cmark 	 & & & \cmark & \cmark \\
Androguard~\cite{desnos2012android}	&	Similarity Comparison	& method-level signature									& 		&	 \cmark  &&& \cmark & \\
\rowcolor{lightgray} Potharaju et al.~\cite{potharaju2012plagiarizing}& Similarity Comparison & abstract syntactic tree		& 		&	 \cmark  &&&\cmark & \\
Ruiz et al.~\cite{ruiz2012understanding} & Similarity Comparison	& count-, set-, sequence-, and relationship-based objects &  &  \cmark  &&&& \\
\rowcolor{lightgray} WuKong~\cite{wang2015wukong}	 &	Similarity Comparison 	& API call sequences	, variable occur times	&  &	 \cmark  &&\cmark & \cmark & \cmark \\
Kim et al.~\cite{kim2015detecting} & Similarity Comparison 	& runtime API invocations	& \cmark  &  & & & \cmark & \\

\rowcolor{lightgray} ResDroid~\cite{shao2014towards} &	Unsupervised Learning	& activities, permissions, intent filters, event handlers, etc. && \cmark  && \cmark & \cmark & \cmark \\
AnDarwin2~\cite{crussell2014andarwin}	& Unsupervised Learning				& program dependency graph			& 	& \cmark 	 &  && \cmark & \cmark \\
\rowcolor{lightgray} AnDarwin~\cite{crussell2013andarwin}	&	Unsupervised Learning				& program dependency graph			& 	& \cmark 	 &  && \cmark & \cmark \\
DroidMat~\cite{wu2012droidmat}		& 	Unsupervised Learning				& permissions, intents, components, API calls, ICC & &	 \cmark 		 &  &&& \\

\rowcolor{lightgray} DAPASA~\cite{fan2017dapasa} & Supervised Learning		& coefficient/distance of sensitive subgraph/motifs	&		& \cmark & \cmark  && \cmark & \cmark \\
DR-Droid2~\cite{tian2017detection} & Supervised Learning		& user interactions, sensitive APIs, permissions			&		& \cmark & \cmark && \cmark & \cmark \\
\rowcolor{lightgray} DR-Droid~\cite{tian2016analysis}		&	Supervised Learning		& user interactions, sensitive APIs, permissions			&		& \cmark & \cmark && \cmark & \cmark \\
DroidLegacy~\cite{deshotels2014droidlegacy}	&	Supervised Learning & frequency of API calls in primary module	&  	& \cmark   & \cmark &\cmark & \cmark & \\
\rowcolor{lightgray} SCSdroid~\cite{lin2013identifying}	&	Supervised Learning			& system call sequences				& 	\cmark	&	 \cmark  & & &  &  \\

PICARD~\cite{aldini2015detection}	&	Runtime Monitoring		& execution trace					& \cmark	 &		 \cmark  &&&  &  \\
\rowcolor{lightgray} DIVILAR~\cite{zhou2014divilar}		&	Runtime Monitoring	&	virtualization-based protection			& \cmark 	& \cmark 	 &&& &  \\
DroidMarking~\cite{ren2014droidmarking} & Runtime Monitoring		& watermarking						&  \cmark	&	 \cmark 	 &&& &  \\	
\rowcolor{lightgray} AppIntegrity~\cite{vidas2013sweetening} & Runtime Monitoring		& package name						&  \cmark &   &&&& \\
AppInk~\cite{zhou2013appink}			&	Runtime Monitoring		&  watermarking		&  \cmark 	& \cmark 	 &&&& \\
\hline
\multicolumn{9}{l}{MS$^\alpha$: Market-Scale, GT$^\beta$: Ground Truth, OR$^\gamma$: Obfuscation Resilience, LN$^\delta$: Library Noise.}
\end{tabular}
}
\end{table*}

\section{Review of State-of-the-art Approaches}
\label{sec:solutions}
The papers collected for the SLR include an approach and experiments related to repackaged apps detection.
We characterise the different approaches and discuss their evaluation scenarios.

\subsection{Taxonomy of approaches}
		\label{subsec:approaches}
Table~\ref{tab:approaches} provides details on categorisation of the different state-of-the-art approaches, 
highlighting the various features each leverages in its proposed approach.
Repackaged app detection can be performed \underline{statically or dynamically}. We also note that there
are static approaches which do not \underline{analyse the bytecode} in the app package for their detection
process. Instead, they solely rely on the resource files accompanying the code.
Various information from apps are leveraged as \underline{features} for identifying repackaged apps. Such
features can be extracted from metadata (e.g., permissions recorded in the Manifest file), 
from the code (e.g., call graphs), or from runtime data (e.g., execution traces).

Based on our review of the \numofpapers{} state-of-the-art studies, we propose a \underline{taxonomy of 5 categories} for the design of state-of-art approaches:

	{\bf \em Similarity computation}-based approaches, developed in 42 out of \numofpapers{} papers, are the most common methodology adopted in the literature. 
	Since Androguard~\cite{androguard, desnos2011android}, which has proposed algorithms for pairwise comparison of apps, 
	several variants using code information (e.g., DNADroid~\cite{crussell2012attack} with dependence graphs, DroidMOSS~\cite{zhou2012detecting} with fuzzy hashing-based fingerprints), layout/resource information (e.g., DroidEagle~   \cite{sun2015droideagle}) or a combination of both (e.g., ResDroid~\cite{shao2014towards}, ViewDroid~\cite{zhang2014viewdroid}) have been developed. Several approaches have further been proposed to improve the scalability of the state-of-the-art in pairwise comparison. Generally,
	these involve a two-step process during which the apps are first pre-processed to extract features that best summarise them. 
PiggyApp~\cite{zhou2013fast} builds vectors using normalised values of extracted features. Thus, instead of computing the similarity between apps based on their code, the authors compute the distance of vectors associated with the apps. Although this state-of-the-art work attempts to escape the scalability problem with pairwise comparisons by relying on the Vantage Point Tree data structure to partition the metric space, it still requires the dataset to contain exhaustively the original apps as well as their repackaged versions. 
Later, Chen et al.~\cite{chen2014achieving,chen2015finding} have proposed to abstract app method code into a single geometric characteristic value, a graph score, to allow even faster comparisons. Their approaches are however also unusable in practice in the context of the myriads of Android markets which are difficult to crawl~\cite{allix2016androzoo} at once so as to have all potential original and repackaged apps in the search space.

	{\bf \em Runtime monitoring}-based approaches, used in 5 approaches, record or/and extract specific information during dynamic execution (or installation) of apps to check whether an app is repackaged or not. Most of those approaches (e.g., AppInk~\cite{zhou2013appink}) aim at repackaging deterrence by providing means for market maintainers to arbitrate/validate whether a watermarked app has been repackaged.

	{\bf \em Supervised learning}-based approaches, implemented in 5 approaches, extract feature vectors from app data and train classifiers that will be used to predict whether an app is repackaged or not. DR-Droid~\cite{tian2016analysis} reuses known features from the malware detection community and applies it to each of the statically identified loosely-coupled parts of a repackaged app. 
	SCSDroid~\cite{lin2013identifying} compares dynamically recorded system call sequences against some pre-learned runtime information of known families for detecting repackaged malware.

	{\bf \em Unsupervised learning}-based approaches, developed in 4 approaches, regroup apps in different sets using advanced learning algorithms with features that can split apps based on their similarity. We differentiate these approaches from simple Similarity computation-based ones, as they radically try to improve the scalability of the pairwise comparison between apps, by focusing on apps that are likely to be repackaged from one another.
	
	{\bf \em Symptom discovery}-based approaches, implemented in only one recent workshop paper, build on the intuitive assumption that repackaging processes leave marks on the repackaged apps. 
	If such marks can be fully characterised, it is possible to spot the symptoms in apps. AndroidSOO~\cite{gonzalez2015exploring} has recently introduced and explored a novel and easily extractable attribute called  String  Offset  Order, which is extracted from string  identifiers  list in  the  {\em classes.dex} bytecode file. Such approaches can normally provide promising results as they can manage to solve most challenges at once, especially the requirement to have the original apps available in the test set, and are unlikely suffering from false positive results if the corresponding symptoms are well-defined.

During SLR paper examination, we have attempted to identify which challenges, among those enumerated in Section~\ref{sec:challenges}, authors
have strived to address. 
In particular, we check that the proposed approach/methodology 
1) meets market constraints (in terms of scalability and usability in practice), 
2) is evaluated based on a constructed reference dataset (whatever its size and representativeness), 
3) explicitly accounts for app obfuscation (to any extent), 
and 4) attempts to reduce the noise of common libraries. 
Details in Table~\ref{tab:approaches} show that no approach addresses all challenges, with Market-scale constraints being the least tackled in the literature.

We further study the most represented similarity computation-based approaches. Details enumerated in Table~\ref{tab:algorithms}
show that the simple Jaccard index is the most shared similarity metric. 
A plethora of approaches are then trying different algorithms for computing the similarity scores.

\begin{table*}[!h]
\centering
\caption{Distance metrics used in Similarity computation approaches.
}
\label{tab:algorithms}
\resizebox{1\linewidth}{!}{
\begin{tabular} { p{0.2\linewidth} c p{0.5\linewidth} c}
\hline
Algorithm	&	Formula$^1$		&	Approaches & Count	\\
\hline
Jaccard		&	$\frac{\vert X \cap Y  \vert}{\vert X \cup Y  \vert}$			& PLayDrone~\cite{viennot2014measurement},	FSquaDRA~\cite{zhauniarovich2014fsquadra}, ViewDroid~\cite{zhang2014viewdroid}, PiggyApp~\cite{zhou2013fast}, JuxtApp~\cite{hanna2012juxtapp}, Wu et al.~\cite{wu2015detect}, Ruiz et al.~\cite{ruiz2012understanding}, Ruiz et al.~\cite{ruiz2014large}, Linares-V{\'a}squez et al.~\cite{linares2014revisiting}, Li et al.~\cite{li2016investigation}, SimiDroid~\cite{li2017simidroid}, APPraiser~\cite{ishii2017appraiser}, Kim et al.~\cite{kim2015detecting}  	 & 13\\[0.3cm]
Euclidean 	&	$\sqrt{\sum_{1}^{n} (x_i - y_i)^2}$			& Potharaju et al.~\cite{potharaju2012plagiarizing}, Soh et al.~\cite{soh2015detecting} & 2\\[0.3cm]
Normalized Compression & $\frac{L_{X \vert Y - min\{L_X,L_Y\}}}{max\{L_X, L_Y\}}$									& Androguard~\cite{desnos2012android}, AndRadar~\cite{lindorfer2014andradar} & 2	\\[0.3cm]
Mahalanobis$^2$	&	$\sqrt{\sum_{1}^{n} \frac{(x_i - y_i)^2}{s_i^2}}$				& RepDetector~\cite{guan2016semantics}	& 1	\\[0.3cm]
Manhattan	&	$\frac{\sum_{1}^{n} \vert x_i - y_i \vert}{\sum_{1}^{n} (x_i + y_i)}$				& WuKong~\cite{wang2015wukong}	 & 1		\\[0.3cm]
Cosine		&	$\frac{\sum_{1}^{n} x_iy_i}{\sqrt{\sum_{1}^{n} x_i^2}\sqrt{\sum_{1}^{n} y_i^2}}$	& Kywe et al.~\cite{kywe2014detecting}, CLANdroid~\cite{linares2016onautomatically}	& 2		\\[0.3cm]
 			& \multicolumn{2}{r}{FUIDroid~\cite{lyu2017efficient}, DroidClone~\cite{alam2016droidclone}, Niu et al.~\cite{niu2016clone}, DroidAnalytics~\cite{zheng2013droidanalytics}, DNADroid~\cite{crussell2012attack}, DroidMOSS~\cite{zhou2012detecting}, ImageStruct~\cite{jiao2015rapid}} &  \\
Customized/Other	& \multicolumn{2}{r}{RepDroid~\cite{yue2017repdroid}, SUIDroid~\cite{lyu2016suidroid}, MassVet~\cite{chen2015finding}, AdRob~\cite{gibler2013adrob}, DroidSim~\cite{sun2014detecting}, DroidEagle~\cite{sun2015droideagle}, DroidKin~\cite{gonzalez2014droidkin}, AndroSimilar2~\cite{faruki2015androsimilar}}	& 21	\\
			& \multicolumn{2}{r}{GroupDroid~\cite{marastoni2017groupdroid}, Chen et al.~\cite{chen2015detecting}, Chen et al.~\cite{chen2014achieving}, FSquaDRA2~\cite{gadyatskaya2016evaluation}, AndroSimilar~\cite{faruki2013androsimilar}, CodeMatch~\cite{glanz2017codematch}}	& \\
\hline
\multicolumn{4}{l}{$^1$$x_i$, $y_i$ are elements of feature set $X$, $Y$.}\\
\multicolumn{4}{l}{$^2$$s_i$ in Mahalanobis algorithm is the standard deviation of $x_i$, $y_i$ over the sample set.}
\end{tabular}
}
\end{table*}

Overall, the literature mostly describes approaches that statically detect repackaged apps by analysing the app bytecode.
Most of the approaches are based on pairwise similarity comparison, which is unfortunately not suitable for market-scale analyses. 
We remind the readers that in this work the scalability issue is referred to the problems of (1) combinatorial explosion and (2) absence of original apps.
Indeed, even when some approaches find relevant features for efficiently speeding the comparison (e.g., Andarwin~\cite{crussell2014andarwin} can analyse an app in 109 seconds), pairwise comparison-based and unsupervised learning-based approaches are still facing the issue of requiring the presence of the original app to find its repackaged versions.
Nevertheless, pairwise similarity comparison is still useful and is often necessary. Actually, in general, the analysis results of any advanced approaches, which may meet market scalability requirements, must still be vetted and confirmed via a pairwise comparison that validates the high similarity score between the suspicious repackaged app and another app.

\begin{tcolorbox}
Most state-of-the-art approaches cannot scale to millions of Android apps. They are thus not practical for market maintainers.
\end{tcolorbox}

\subsection{Review of Evaluation Setups and Artefacts}		
\label{subsec:evaluations}

Authors of state-of-the-art approaches have argued that their proposed features enumerated in Table~\ref{tab:approaches} 
are effective for identifying repackaged apps. However, in the absence of a comprehensive comparative assessment of existing approaches, one question remains open for the community:  {\em what is the minimal feature set that is most effective in discriminating repackaged apps from non-repackaged ones? }

We survey the origin of datasets used for experiments, their sizes, and
the availability of tools and data from state-of-the-art approaches for use by other researchers.
Table~\ref{tab:evaluations} provides the details of this assessment information for the reviewed
publications.

\begin{table}[!t]
\centering
\caption{Summary of Examined Approaches based on Their Evaluation Metrics.}
\label{tab:evaluations}
\resizebox{1.0\linewidth}{!}{
\begin{tabular} { r c c c 	c}
\hline
Publication & Tool & Dataset & \#. of  	 		& Genome \\
			 & Available & Available &  Apps  & \\ 
\hline
CodeMatch~\cite{glanz2017codematch} & \cmark  & \cmark  & (10000,100000) & \\
\rowcolor{lightgray} DR-Droid2~\cite{tian2017detection} & & & (1000,10000) & \cmark \\
DAPASA~\cite{fan2017dapasa} & & & (10000,100000) & \cmark \\
\rowcolor{lightgray} FUIDroid~\cite{lyu2017efficient} & & & (10000,100000) & \\
APPraiser~\cite{ishii2017appraiser} & & & (1000000, $\infty$) & \\
\rowcolor{lightgray} RepDroid~\cite{yue2017repdroid} & \cmark & \cmark & (100,1000) & \\
SimiDroid~\cite{li2017simidroid} & \cmark & \cmark & (1000,10000)  & \\
\rowcolor{lightgray} GroupDroid~\cite{marastoni2017groupdroid} & & & (1000,10000) & \cmark \\

CLANdroid~\cite{linares2016onautomatically} & \cmark & \cmark   & (10000,100000) & \\
\rowcolor{lightgray} DR-Droid~\cite{tian2016analysis}							&	&&	(1000,10000)  		& \cmark\\

DroidClone~\cite{alam2016droidclone} &	&	& (100,1000) & \\

\rowcolor{lightgray} FSquaDRA2~\cite{gadyatskaya2016evaluation}    & \cmark	&	& (1000,10000) & \\

Li et al.~\cite{li2016investigation}						&	&\cmark$^\alpha$	  & (1000000, $\infty$) & \\

\rowcolor{lightgray}  Niu et al.~\cite{niu2016clone}    &	&	& - & \\

RepDetector~\cite{guan2016semantics}						&	&&	(1000,10000)	  	& \\

\rowcolor{lightgray} SUIDroid~\cite{lyu2016suidroid}    &	& 	& (100000,1000000) & \\

Kim et al.~\cite{kim2015detecting} & & & (100,1000) & \\

\rowcolor{lightgray} AndroidSOO~\cite{gonzalez2015exploring}					&	&&	(10000,100000)		& \cmark	 \\

AndroSimilar2~\cite{faruki2015androsimilar} &	&	& (10000,100000) & \cmark \\

\rowcolor{lightgray} Chen et al.~\cite{chen2015detecting}						&	&&	(1000,10000)		& \cmark \\

DroidEagle~\cite{sun2015droideagle}						&	&&	(1000000, $\infty$)	 	& \\
\rowcolor{lightgray} ImageStruct~\cite{jiao2015rapid}							&	&&	(10000,100000)		& \cmark \\

MassVet~\cite{chen2015finding}							&	&&	(1000000, $\infty$) 	& \\

\rowcolor{lightgray} PICARD~\cite{aldini2015detection}						&	&&	(0,100)		& \\

Soh et al.~\cite{soh2015detecting}						&	&&	(100,1000)		& \\

\rowcolor{lightgray} Wu et al.~\cite{wu2015detect}							&	&&	(1000,10000) 		&  \\

WuKong~\cite{wang2015wukong}								&	&&	(100000,1000000)		& \\
\rowcolor{lightgray} AnDarwin2~\cite{crussell2014andarwin}					&	&&	(100000,1000000)		& \\
AndRadar~\cite{lindorfer2014andradar}					&	&&	(100000,1000000) 	& \cmark \\

\rowcolor{lightgray} Chen et al.~\cite{chen2014achieving}						&	&&	(10000,100000)		& \\
DIVILAR~\cite{zhou2014divilar}							&	&&	(0,100)	 	& \\

\rowcolor{lightgray} DroidKin~\cite{gonzalez2014droidkin}						&	&&	(1000,10000) 		& \cmark	\\
DroidLegacy~\cite{deshotels2014droidlegacy}				&	&&	(1000,10000)		& \cmark \\

\rowcolor{lightgray} DroidMarking~\cite{ren2014droidmarking} 					& 	&&	(100,1000)		& \\

DroidSim~\cite{sun2014detecting}							&	&&	(100,1000)	& \cmark \\

\rowcolor{lightgray} FSquaDRA~\cite{zhauniarovich2014fsquadra}				& \cmark	&&	(10000,100000)	& \\

Kywe et al.~\cite{kywe2014detecting}						&	&&	(10000,100000)	& \\
\rowcolor{lightgray} Linares-V{\'a}squez et al.~\cite{linares2014revisiting}	& 	&	\cmark$^\alpha$ &	(10000,100000)		& \\

PlayDrone~\cite{viennot2014measurement}					&  	& \cmark$^\alpha$ 	&	(1000000, $\infty$)	& \\

\rowcolor{lightgray} ResDroid~\cite{shao2014towards}							&	&&	(100000,1000000)		& \\

Ruiz et al.~\cite{ruiz2014large}							&	&&	(100000,1000000)		& \\

\rowcolor{lightgray} ViewDroid~\cite{zhang2014viewdroid}						&	&&	(10000,100000)		& \\

AdRob~\cite{gibler2013adrob}								&	&&	(100000,1000000)		& \\
\rowcolor{lightgray} AnDarwin~\cite{crussell2013andarwin}						&	&&	(100000,1000000)		& \\

AndroSimilar~\cite{faruki2013androsimilar} &	& & (10000,100000) & \cmark \\
\rowcolor{lightgray} AppInk~\cite{zhou2013appink}								&	&&	(0,100)	& \\
AppIntegrity~\cite{vidas2013sweetening} 					& 	&&	(10000,100000)	& \\
\rowcolor{lightgray}  DroidAnalytics~\cite{zheng2013droidanalytics} 			& 	& \cmark$^\alpha$ &	(100000,1000000)		& \\
PiggyApp~\cite{zhou2013fast}								&	&&	(10000,100000)		& \\
\rowcolor{lightgray} SCSdroid~\cite{lin2013identifying}						&	&&	(100,1000)		& \\
Androguard~\cite{desnos2012android}						& \cmark	& & -	& \\
\rowcolor{lightgray} DNADroid~\cite{crussell2012attack}						&	&&	(10000,100000)		& \\
DroidMat~\cite{wu2012droidmat}							& 	&&	(1000,10000)	& \\
\rowcolor{lightgray} DroidMOSS~\cite{zhou2012detecting}						&	&&	(10000,100000)		& \\
JuxtApp~\cite{hanna2012juxtapp}							&	&&	(100000,1000000)	& \\
\rowcolor{lightgray} Potharaju et al.~\cite{potharaju2012plagiarizing}			& 	&&	(1000,10000)	& \\
Ruiz et al.~\cite{ruiz2012understanding} 				& 	& \cmark$^\alpha$ &	(1000,10000)		& \\

\hline
\multicolumn{5}{l}{\cmark$^\alpha$: dataset without repackaging labels}
\end{tabular}
}
\end{table}

\textbf{Tool availability} Among the \numofpapers{} publications proposing approaches for detecting
repackaged apps, only 7 have made their tool support available.

\textbf{Datasets availability} Only 4 approaches have publicly released a ground truth of similarities among Android apps. 
Five (5) approaches have released the original set of apps where they searched for repackaged apps.

\textbf{Dataset size} There is a huge variation among the sizes of datasets used in
the evaluation setups of state-of-the-art approaches. Five studies have gone over the one-million apps mark, 10 studies have analysed more than 100 thousand apps (although less than 1 million), 30 studies have analysed between 1000 and 100,000 apps, 
7 papers have analysed between 100 and 1000 apps.
Three papers have even been assessed on less than 100 apps. Given the size of the official market alone, there is room to improve the scale of the experiments performed in the literature.

\textbf{Dataset diversity} We also checked the origin of the datasets and found that many approaches
collect their experimental datasets from 1 or few sources. We note that those approaches that used several sources are not necessarily the ones that assessed on the largest datasets.

\begin{figure*}[!h]
    \centering 
    \includegraphics[width=0.8\linewidth]{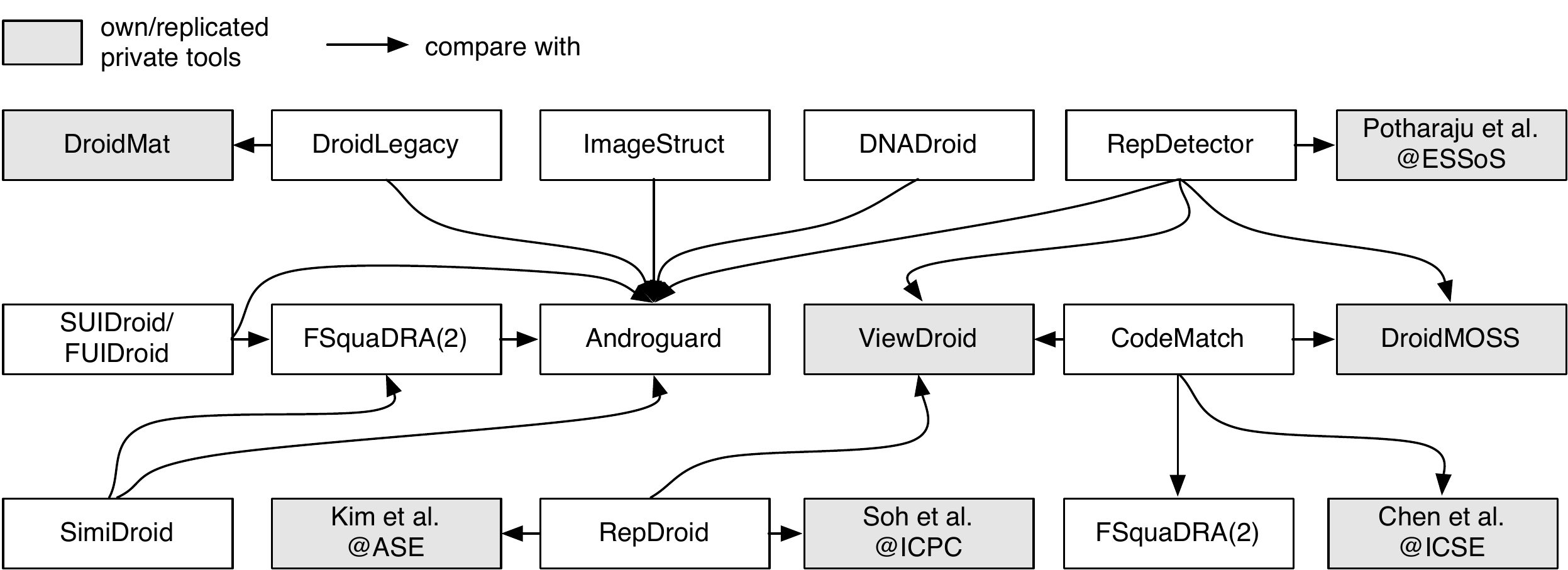}
    \caption{Relationship of comparison among state-of-the-art approaches.}
    \label{fig:comparison}
\end{figure*}

We have further investigated to what extent state-of-the-art approaches have been compared in
the literature. Given the lack of data sharing and the unavailability of tool support from competitor approaches, little comparative evaluation has been presented in the literature. 
Among the \numofpapers{} publications, only nine (9) have performed a comparative study 
against the similarity scores of other approaches (e.g., the Androguard~\cite{androguard} tool).
Fig.~\ref{fig:comparison} summarises the graph of comparison among the different state-of-the-art approaches. We note that other comparative assessments are performed on authors' previous studies (which, by the way, are not available to others)
or by replicating, to the best of their effort, some basic similarity computation-based approach.

Finally, we investigate how the accuracy of repackaged app detection is evaluated in state-of-the-art literature.
In the absence of an external ground truth, authors apply their approach on random datasets and then manually verify the
findings, on a sampled subset, to compute efficiency, leading to the introduction of potential researcher bias. For evaluating their recent centroid-based approach, Chen et al.~\cite{chen2014achieving} have randomly selected and checked 359 apps among thousands of apps that are flagged as cloned and found that their approach has zero false positive.
A few approaches (7, cf. Fig.~\ref{fig:comparison}) use Androguard as a proxy to check the similarity of the detected repackaging pairs, adding some confidence to their evaluation setup. 
Supervised Learning approaches rely on the Genome dataset or other malware from VirusTotal to build training and test sets.

\begin{tcolorbox}
Non-disclosure of tools and datasets is leading to redundant research and does not encourage innovation since 
there is  limited opportunities to reproduce, validate and compare.

\end{tcolorbox}

\section{Dataset Construction}
\label{sec:future_directions}

State-of-the-art work in the literature claims high-performance rates for their proposed approaches.
Unfortunately, their shortcomings in opening their datasets and implementations to comparative assessment by other researchers are actually blocking further research into the problem. 
Since Android app repackaging remains relevant today\footnote{As demonstrated in the Lookout 
blog~\cite{pokemon_go}, which details their analysis of the various repackaging versions of the popular Pokemon Go app (repackaging with a trojan included, repackaging for cheating, repackaging with adware included, etc.)},
we propose to reboot this research topic, in the hope of 
encouraging novel contributions that will tackle efficiently the different challenges enumerated in Section~\ref{sec:challenges}.

To that end, we propose to {\bf build an extensive dataset}, namely \repo{}, for assessing repackaged detection algorithms. 
Such a dataset includes a set of repackaged apps accompanied with a ``proof'' of their repackaged state by providing the original apps with which they form repackaging pairs, following the same definition: A repackaged app pair (1) has at least 80\% of code similarity between its two apps and (2) has its two apps are signed by different developers.
Our work builds upon the popular AndroZoo repository, which can serve as an exchange repository for describing one's dataset using the hash values of apps.

\begin{figure*}[!th]
    \centering 
    \includegraphics[width=\linewidth]{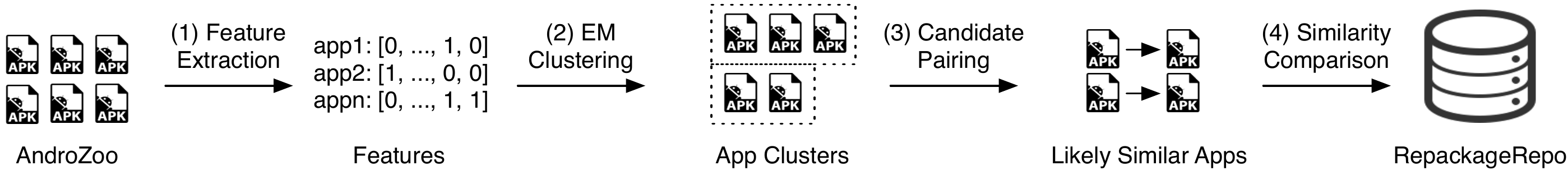}
    \caption{Overview of \repo{} Construction. Step (1), (2), and (3) for Splitting the Search Space. Step (4) for Fast, Approximate, Brute-force Similarity Comparison.}
    \label{fig:overview_dataset}
\end{figure*}

The collection of \repo{} is done in a systematic way.
Fig.~\ref{fig:overview_dataset} illustrates an overview of the construction process. We leverage the AndroZoo dataset~\cite{allix2016androzoo}, which (by the time of this study) includes over 5 million apps continuously crawled from 13 markets including the official Google Play and several alternatives markets such as AppChina, as well as online repositories such as F-Droid and the Genome project. 
To find repackaging pairs, we take the traditional, time-consuming, but most accurate, approach of performing similarity computations. 
To optimize the process, we devise
a two-phase approach for identifying repackaging pairs. 
We now detail these two phases separately.

\subsection{Splitting the search space} 
\label{subsec:split}

Because of time and computing resources constraints, it is virtually infeasible to perform pairwise comparisons for all possible app combinations in a dataset such as AndroZoo. Instead,
we propose to rely on a clustering-based approach to split the search space, so as to focus on comparing only likely similar apps. 
As illustrated in Fig.~\ref{fig:overview_dataset}, this phase is actually made up of three steps:

\begin{itemize}

\item Step (1): Feature Extraction.
We abstract each app into a representative feature vector. To ensure processing speed, we focus
on features that are easily extractable from an APK file. 
Those include class names, declared permissions, declared actions and intent-filters. 
These features will prevent for example from regrouping a game app with a messenger app and will ensure that similar apps are included together in the same cluster. 

\item Step (2): EM Clustering.
We leverage the Expectation Maximization (EM) algorithm~\cite{moon1996expectation} to regroup the AndroZoo apps into different clusters. 
EM is preferred to other popular clustering algorithms, such as K-mean, because it does not require to be parameterised with the number of clusters that should be produced.

\item Step (3): Candidate Pairing.
At the end of this phase, we consider the set of apps in each cluster and form candidate combinations of repackaging pairs.
Given two apps in a cluster, the candidate pair is formed by considering the one created before (based on the creation time of the DEX file) as the original app while the remaining one as the repackaged version.
Because we consider repackaging to be mostly a parasite activity, we drop candidate pairs where the apps have been signed with the same developer certificate.
Indeed, apps that are signed by the same certificate are usually considered to be app variants of a same company or app versions of the same app, which are unlikely to be repackaged versions.

\end{itemize}

\subsection{Fast, approximate, brute-force similarity comparison}
Given a candidate pair of apps (app$_1$, app$_2$) found in a cluster, we compute four similarity metrics, and calculate a score based on Formula~(\ref{eq:similarity}).
\begin{itemize}
	\item {\em identical} ($I$), which represents the number of methods (signature + body) which are shared by both apps; 
	\item {\em similar} ($S$), representing the number of similar (same signature but different body contents) methods between two apps;
	\item {\em new} ($N$), representing the number of new methods that were added in app$_2$ in comparison with app$_1$; and
	\item {\em deleted} ($D$), representing the number of such methods that exist in app$_1$ but not in app$_2$.
\end{itemize}

\begin{equation}\label{eq:similarity}
\footnotesize
similarityScore := max\{\frac{|I|}{|I|+|S|+|D|}, \frac{|I|}{|I|+|S|+|N|}\}
\end{equation}

To ensure a fast computation of the above metrics, we use an approximative representation of 
app method contents by mapping the different statement types to alphabet characters. 
For example, the simplified code snippet presented in Listing~\ref{code:snippet} could be represented by string $acc$, where an interface and virtual invocation statement are respectively mapped to character $a$ and $c$. 
All variables are thus dropped
from the comparison. The contents of the different methods thus go through a 
code-to-text transformation leading to short strings for which efficient similarity 
analysis algorithms exist. By ignoring easily-manipulable variable names, our code-to-text
 transformation enables the brute-force comparison to be resilient to simple obfuscations which
 are commonly performed during repackaging.
Although we have attempted to fasten the pairwise comparisons, 
our  similarity analysis still takes roughly one month to finish on all the candidate pairs.

\begin{lstlisting}[
caption={Simplified Code Snippet of Android Apps. 
This Code Snippet is Presented at the Jimple Level, where Our Similarity Analysis is Implemented on top of Soot, in which Jimple is the Default Intermediate Representation (IR) Code.},
label=code:snippet, float=t]
$r2 = interfaceinvoke $r1.<WindowManager: Display getDefaultDisplay()>();
virtualinvoke $r2.<Display: int getWidth()>();
virtualinvoke $r2.<Display: int getHeight()>();
\end{lstlisting}

Finally, after confirming the effectiveness of our pairwise comparison methodology, we further set the similarity threshold to 80\% to decide that a pair of apps is a repackaging pair. We aim to be conservative by selecting a threshold that is more strict than those used in the literature~\cite{zhauniarovich2014fsquadra, crussell2012attack}.
We have performed experiments to validate that the approximation in Phase 2 of the approach is preserving similarity scores, which basically are in line with the scores computed by comparing the exact statements.
Furthermore, we also validate that the scores that we obtain is similar to those obtained with state-of-the-art tools such as AndroGuard, which is basically in agreement with our collected pairs (more details are given in Section~\ref{subsec:replication}).

\subsection{Overall Results}

Based on the AndroZoo dataset, we are able to collect 15,297 repackaging pairs to be shared as repackaging reference dataset.
We find that many apps are repackaged several times by different attackers, with a minimal, mean, and maximum times of 1, 2.168, and 176, respectively.
Overall, our \repo{} dataset includes 15,297 repackaged apps for 2,776 original apps.
Table~\ref{tab:top3original} shows the top three original apps that are repackaged by over 100 different attackers.
Interestingly, all those three apps are from the official Google Play store, suggesting that Google Play apps are somehow more favoured by attackers to repackage and distribute. 

\begin{table*}[!th]
\centering
\caption{Top Three Original Apps that have been Repackaged by Over 100 Different Attackers.} 
\label{tab:top3original}
\scriptsize
\resizebox{\linewidth}{!}{
\begin{tabular}{ c c c c}
\hline
SHA256 (Original App) & Package Name & Market & Repackaged Certs \\ \hline
9CC2EAEF8636AE77794ACDF085A2C241A98E620581391D41FBC5D39D69528E53 & com.algorythmicstudios.droid & play.google.com & 176 \\
34084F29D69F2056E776B1F6BA3B1174D07C192F4EF2AF7CE793B0DE97C517C9 & com.theindievelopers.stacktothirty & play.google.com & 109 \\
D178AA7FC82311AF6536ECD7872FAEC9C1111E233EF25798F1E157F375862FCC & com.HatchWorks.BabyDiscoverAquatic & play.google.com & 107 \\
\hline
\end{tabular}
}
\end{table*}

\begin{figure}[!th]
    \centering 
    \includegraphics[width=0.7\linewidth]{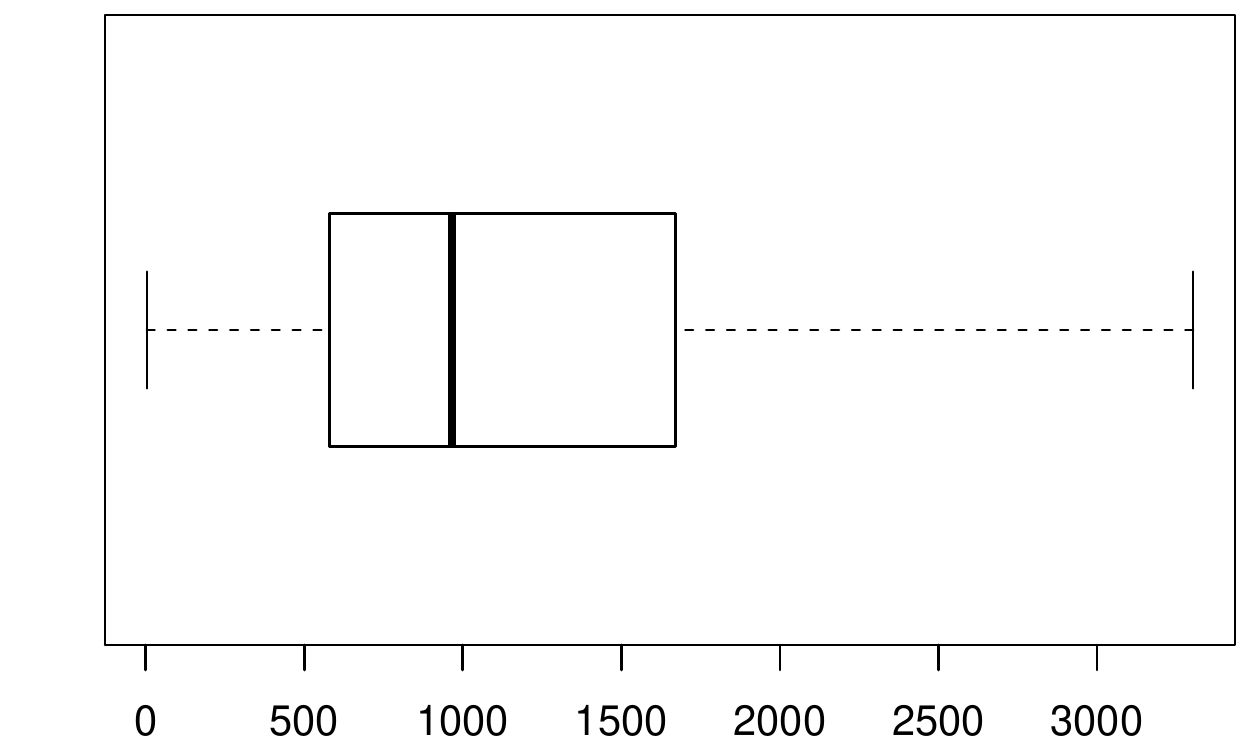}
    \caption{Distribution of DEX Size of \repo{} Apps (in KB).}
    \label{fig:dist_size}
\end{figure}

Fig.~\ref{fig:dist_size} plots the distribution of DEX size of all the collected \repo{} apps, 
where the size ranges from 3.67 KB (minimal) to 16,180 KB (maximum), with a median and mean size of 965.2 KB and 88.67 KB respectively.
This distribution suggests that \repo{}  is quite diverse, containing small-size, middle-size, and large-size Android apps.
We then go one step deeper to investigate the changes of DEX size between the two apps of a given pair.
Among the 15,297 app pairs, over 70\% of them have shown that repackaging will eventually enlarge the DEX file, suggesting additional code is usually injected during repackaging.
However, for nearly 30\% of repackaging cases, the DEX size of repackaged apps are smaller than that of the original apps.

\begin{figure}[!th]
    \centering 
    \includegraphics[width=0.7\linewidth]{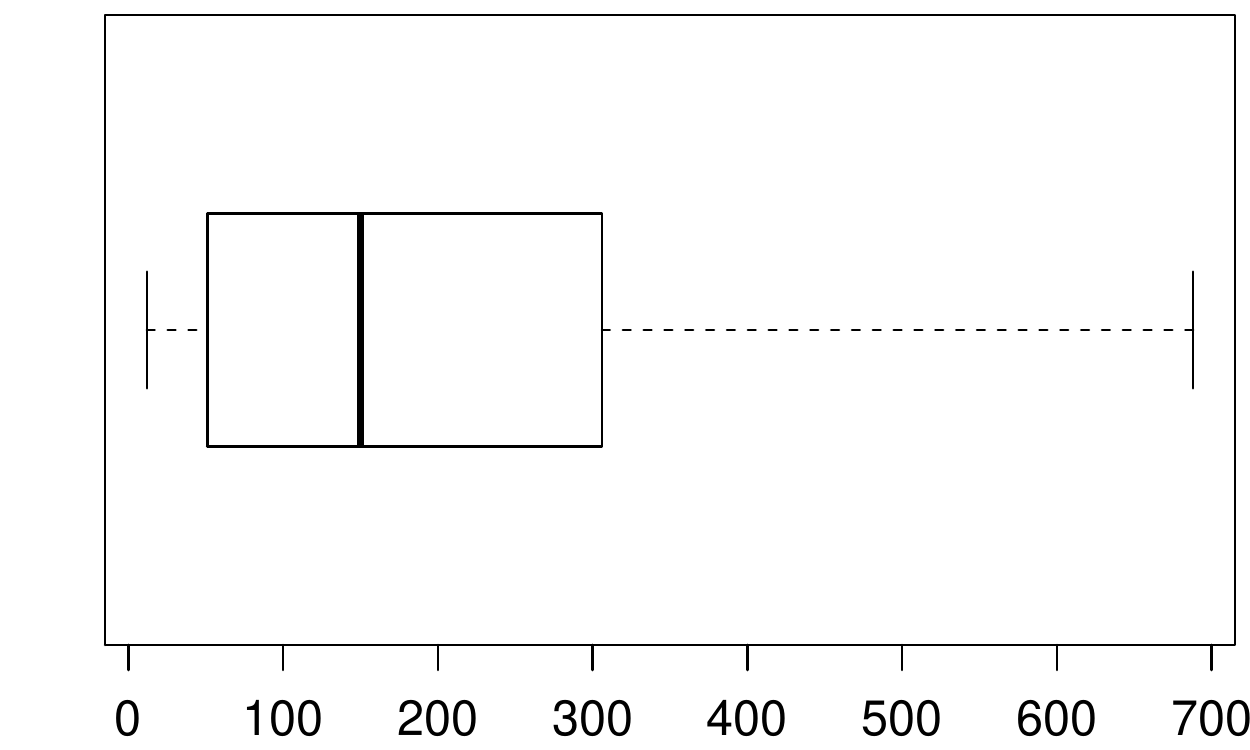}
    \caption{Distribution of the Number of Resource Files.}
    \label{fig:dist_res}
\end{figure}

\begin{figure}[!th]
    \centering 
    \includegraphics[width=0.7\linewidth]{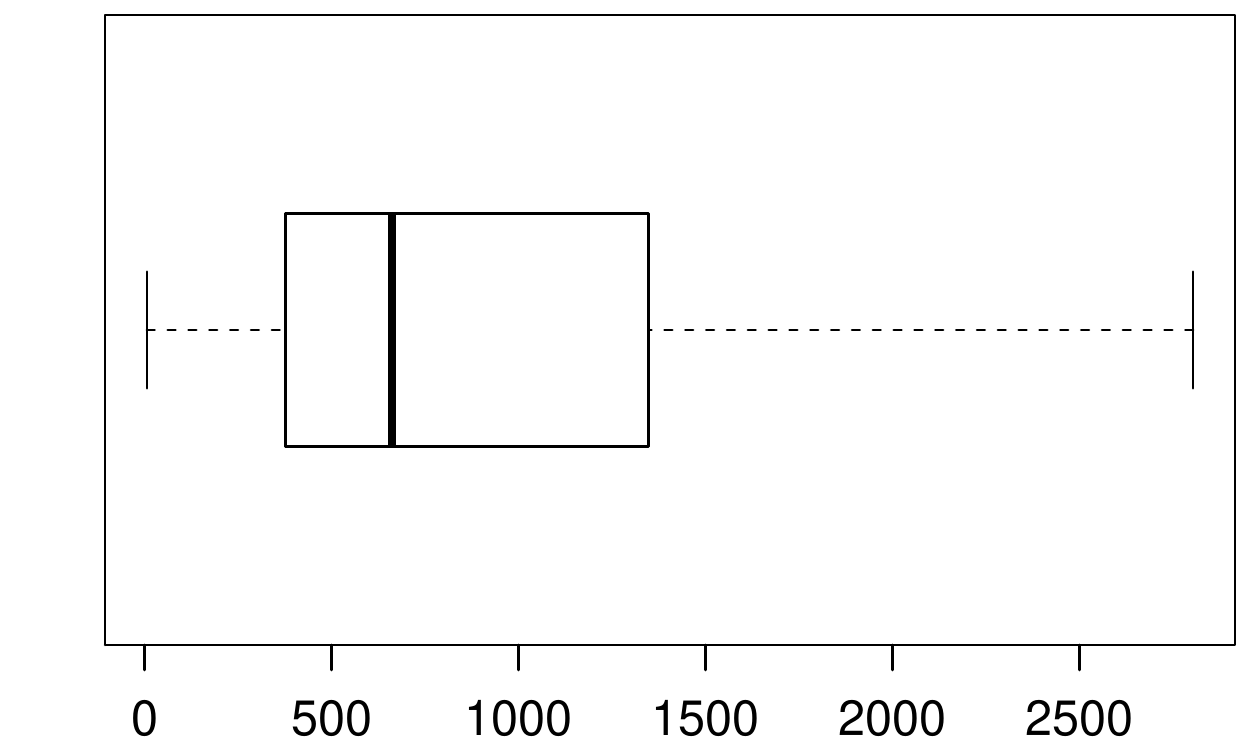}
    \caption{Distribution of the Number of Java Files.}
    \label{fig:dist_smali}
\end{figure}

Fig.~\ref{fig:dist_res} and Fig.~\ref{fig:dist_smali} further respectively plots the distribution of the number of Resource Files and Java Files for the benchmark apps.
The number of resource files ranges from 12 to 10,529\footnote{The maximum number is considered as an outlier so that it is not presented at the boxplot. This explanation also applies to other boxplots.}, where the median and mean values are 150 and 247, respectively.
For Java files, the number ranges from 6 to 7225, where the median and mean values are 662 and 1050, respectively.
These two distributions once again suggest that the RePack dataset is quite diverse, where both apps with a small number of resource/Java files and with a large number of resource/Java files are included.
Interestedly, Spearman's rank correlation coefficient (i.e., $\rho < 0.35)$ suggests that there is no strong correlation between the number of resource files and the number of Java files, further confirming the diversity of our benchmark dataset.

\begin{figure}[!th]
    \centering 
    \includegraphics[width=0.7\linewidth]{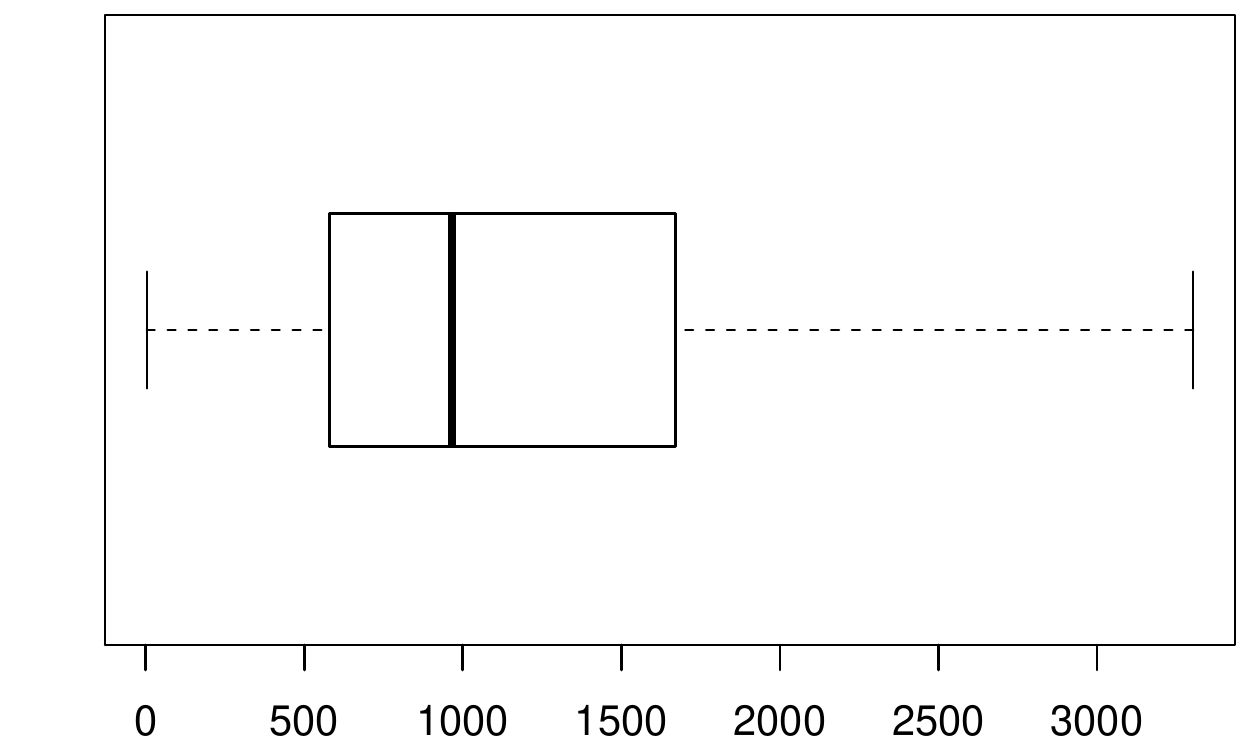}
    \caption{Distribution of Creation Time Diff between \repo{} Pairs (in Day).}
    \label{fig:dist_time_diff}
\end{figure}

Fig.~\ref{fig:dist_time_diff} plots the distribution based on the diff of Creation time between the two apps of a given repackaging pair.
Comparing to the creation time of the original apps, the repackaging delay ranges from several days to several years with an average, 88.67 days.
This distribution also suggests that our collected \repo{} apps are diverse, containing different repackaging cases that could be interesting to malware analysts to investigate.

\begin{table*}[!h]
\centering
\caption{Manual validation results of randomly selected repackaged app pairs (five examples).}
\label{tab:validation}
\resizebox{\linewidth}{!}{
\begin{tabular} { l | c c | c c c c}
\hline
Repackaged  & Code & Resource & \multicolumn{4}{c}{Manual Observation (Same)}\\
App Pair	&	 Similarity & Similarity	& Package Name &	App Name & Icon & Main UI Skeleton	\\
CCAC0E/3CDCEF  & 0.990 & 0.982 & \xmark & \xmark	& \xmark	& \cmark \\
AE064D/6F2081 	 & 0.905 & 0.357 & \xmark &  \xmark 	& \xmark	& \cmark \\
CA9ABD/78940D  & 0.987 & 0.982 & \cmark & \cmark	& \cmark	& \cmark \\
207372/707ED8  & 0.866 & 0.929 & \xmark & \xmark	& \xmark & \cmark \\
FF57A0/74E764  & 0.995 & 0.776 & \xmark &  \xmark & \xmark & \cmark \\

\hline
\end{tabular}
}
\end{table*}

Finally, we use the AVClass~\cite{sebastian2016avclass} labelling tool to assess the diversity of
repackaged malicious apps in our dataset. 
Given a repackaged app, we consider it as malicious as long as one of VirusTotal hosted anti-virus products flags it as such.
For a given app and its labels from VirusTotal anti-virus engines, AVClass outputs a unique name of malware or adware family. 
We feed AVClass with all the repackaged malicious apps in our dataset.
AVClass has successfully classified 5,960 repackaged apps of our dataset into 45 known families, while the Genome dataset includes 49 families. 
Moreover, our dataset includes about 9,337 repackaged apps that AVClass is not able to categorize into a known malware or adware family.

Overall, all the aforementioned studies, covering different aspects, suggest that our collected repackaged pairs, namely \repo{}, is quite diverse, and therefore is reliable to be leveraged to support various analyses.
Last but not the least, as discussed in our previous work~\cite{li2016investigation}, for detecting repackaged Android apps, common libraries may cause both false positive and false negative results.
Hence, we extend our benchmark to also provide library information for each repackaged pairs.
We hope the extended information can encourage the community to innovate in-depth analyses for better understanding the facts between libraries and app clones, including malicious ones.
The library usage (i.e., the 1,113 libraries summarised by Li et al.~\cite{li2016investigation}) of each app in the benchmark has been also made publicly available in our replication dataset.
Moreover, to facilitate the use of library information for Android-based analyses, we provide a research-based prototype tool called LibExcluder for generating library-free versions of given Android apps.
The goal of LibExcluder is to remove library code from a given Android app.
Given a whitelist of common libraries, LibExcluder takes as input an Android app and outputs a new app version, which is generally as same as the inputted one except that the code belonging to libraries configured in the whitelist are excluded.
Therefore, LibExcluder presents to existing state-of-the-art approaches a new app version where library code no longer exists. 
Without any modification (i.e., being non-invasive), existing approaches such as IccTA~\cite{li2015iccta} can benefit from this work to perform library-free analyses.

\begin{tcolorbox}
Our dataset is, to the best of our knowledge, the largest one containing repackaged app pairs.
It is built from a representative set of apps, and includes a diverse set of repackaged apps.
\end{tcolorbox}

\subsection{Manual Validation of Random Samples}

\begin{figure}[!t]
    \centering
	\begin{subfigure}[b]{0.45\linewidth}
		\includegraphics[width=\linewidth]{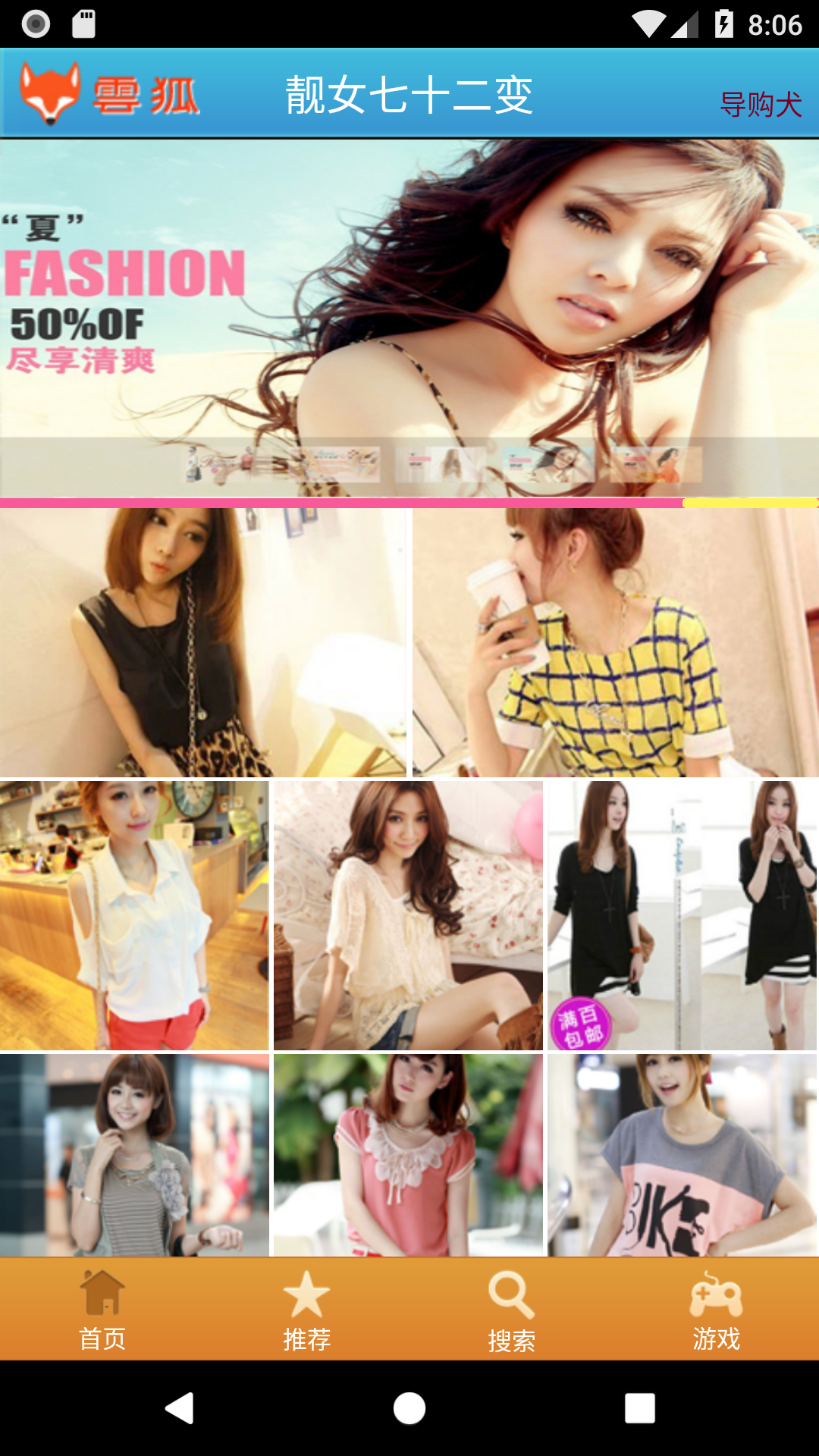}
		\caption{Original App.}
		\label{fig:app1}
	\end{subfigure}\ \ \ \ %
	\begin{subfigure}[b]{0.45\linewidth}
		
		\includegraphics[width=\linewidth]{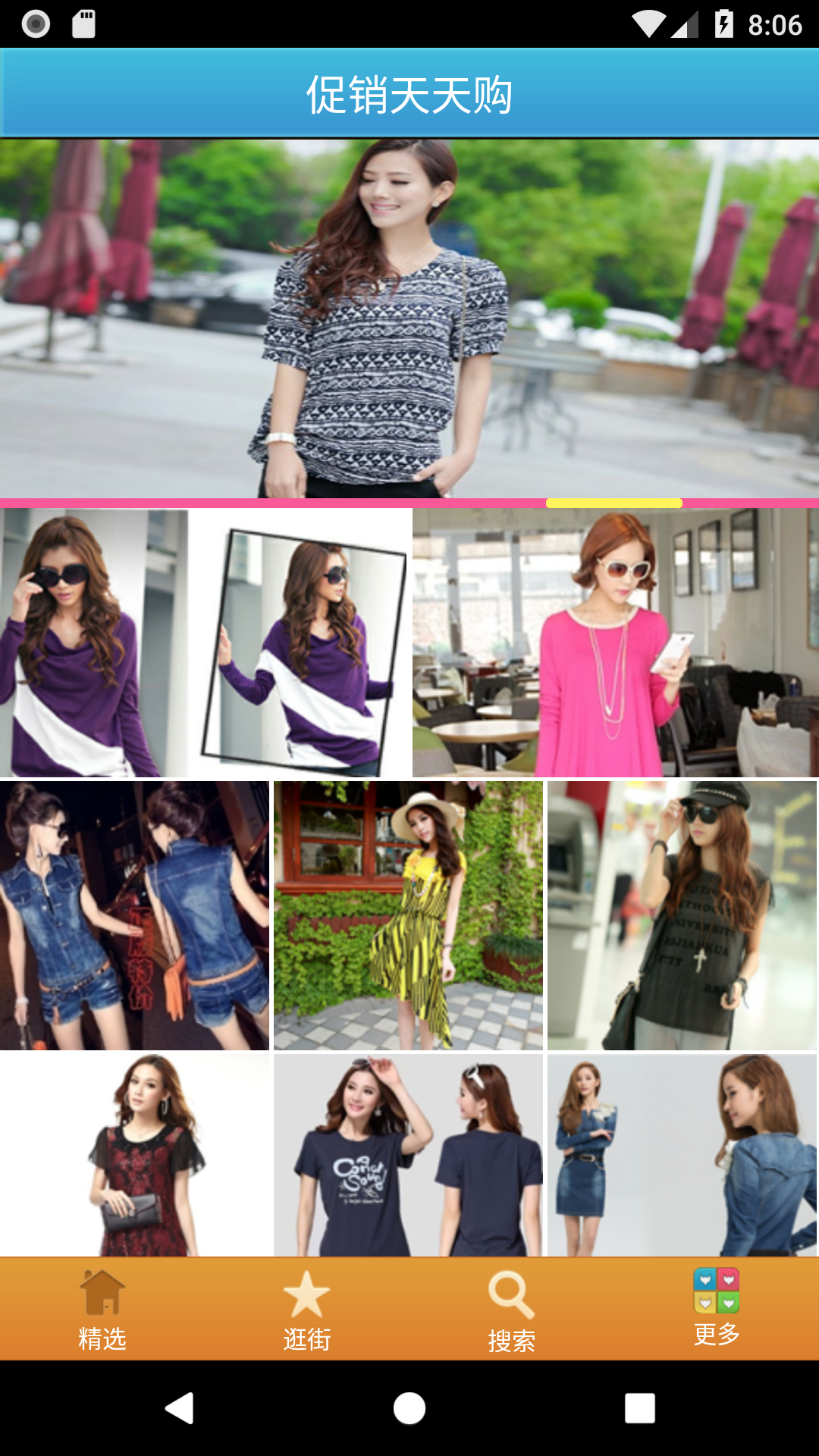}
		\caption{Repackaged App.}
		\label{fig:app2}
	\end{subfigure}
    
	\caption{The first page of apps FF57A0 (left) and 74E764 (right), which more or less share the same UI Skeleton.} 
    \label{fig:demo_example}
\end{figure}

One of the major goals of constructing a benchmark of Repackaged Android apps is to support replication and comparison studies by the community.
Towards demonstrating this ability, we need to ensure in the first place that our constructed benchmark is reliable.
To this end, we randomly select 100 pairs of apps and manual validate their similarities.
Based on the Sample Size Calculator provided by Creative Research Systems\footnote{https://www.surveysystem.com/sscalc.htm}, it is representative to sample100 pairs out of around 16,000 pairs, with a confidence level at 95\% and a confidence interval at 10.
It is actually non-trivial to decide whether a given two apps (which share over 90\% of the code and are signed by different developers) are repackaged from one to another manually.
We hence resort to dynamic analysis to validate the selected pairs.
Given a pair of apps, we manually install them respectively on two emulators set up with exactly the same configurations.
After the apps are installed, we manually launch and play with them and observe the similarity and difference between the two apps.
Among the 100 selected pairs, our manual validation confirms that 89 of them are clearly repackaged pairs (sharing exactly the same UI page or at least similar UI skeleton), giving an accuracy of 89\% at least for our harvested benchmark.
The remaining 11 pairs have big changes in their UI pages that cannot be soundly confirmed.
Nonetheless, we remind the readers that those app pairs, despite having different UI pages, have shared over 80\% of code and thus could be repackaged app pairs as well.
Table~\ref{tab:validation} illustrates five samples of the details we have observed from our manual validation process.
The first column presents the hash values\footnote{SHA256s, only the first six letters are shown.} of the selected pairs.
The second and third columns illustrate the similarity scores yielded by SimiDroid, based on its method-based and resource-based similarity analyses, respectively.
The last four columns show whether the package name, app name, icon and the main UI skeleton are respectively the same between the two apps in a pair.
Regarding the main UI skeleton, we consider it is the same as long as the layout is more or less the same.
For example, in this work, we consider the two pages shown in Fig.~\ref{fig:demo_example} (collected from pair \emph{FF57A0/74E764}) have the same UI skeleton.
If two apps (1) have over 90\% of code similar from one another, (2) are signed by different developers (or teams), and (3) have similar look and feel (i.e., similar UI skeleton), we consider these two apps as true repackaged app pair.
Hence, our manual validation confirms that all the randomly selected app pairs from our benchmark are true repackaged app pairs, suggesting that our benchmark is quite reliable.
Subsequently, our benchmark should be capable of supporting replication and comparison studies between state-of-the-art approaches.

Table~\ref{tab:validation} further reveals that a repackaged app may (or may not) change the package name, app name and icon of the original app.
These findings once again suggest that our benchmark is diverse and is representative to different types of repackaging cases.

\subsection{Supporting Replication and Comparison of State-of-the-art work}
\label{subsec:replication}

Towards demonstrating the ability to support replication and comparison studies, we revisit a couple of literature approaches for repackaging detection which have made their associated tools available.
As shown in Table~\ref{tab:evaluations}, only seven research approaches have made some tools available. 
Nevertheless, not all of them are applicable for our study: CLANdroid, CodeMatch and RepDroid cannot be directly executed as they former two approaches require a corpus preprocessing step and while the last one expects a complicated runtime environment with hard-coded platform dependencies.
FSquaDRA2 and FSquaDRA share the same basics in their approaches, experimenting one of them should be enough.
As a result, we focus on replicating the experiments of FSquaDRA (resource-based comparison), Androguard (approximate code-based comparison), and SimiDroid (exact code-based comparison) based on our \repo{} dataset.

\begin{figure}[!th]
    \centering 
    \includegraphics[width=0.8\linewidth]{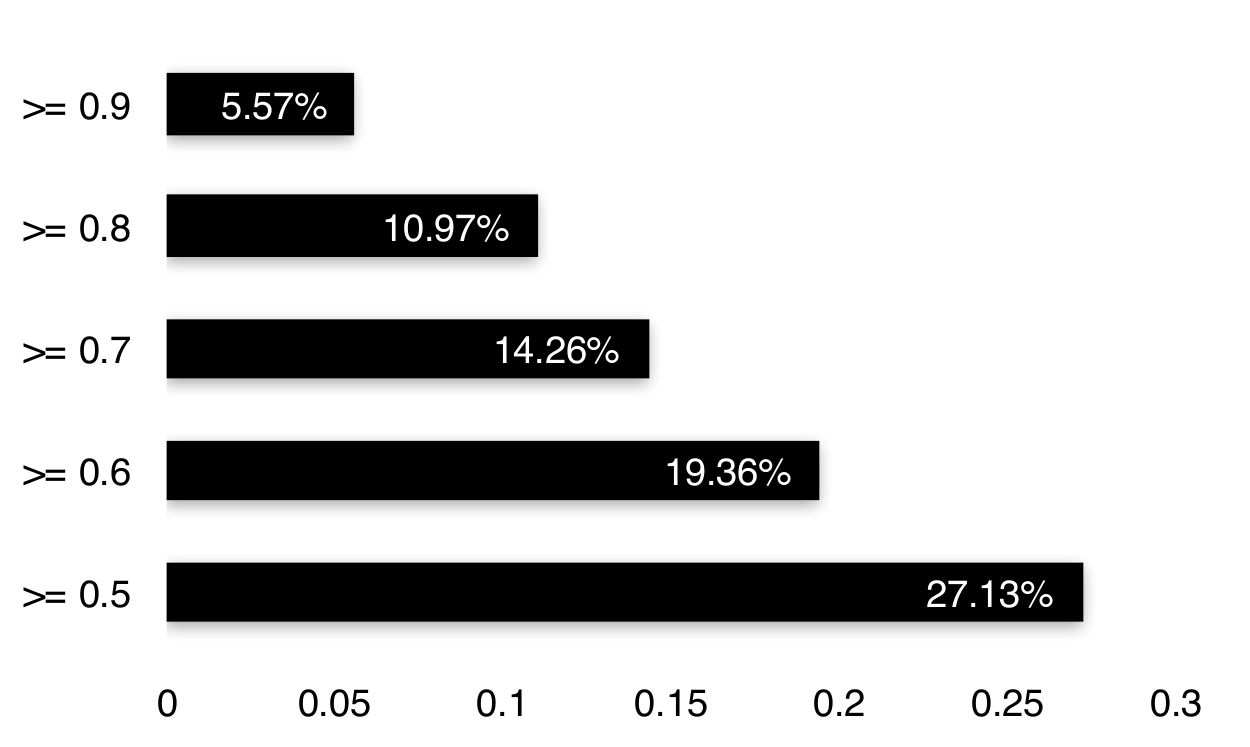}
    \caption{FSquaDRA Results by Different Thresholds.}
    \label{fig:fsquadra_thresholds}
\end{figure}

In this work, we use \emph{recall} to characterise the ability of state-of-the-art tools to detect repackaged apps\footnote{The reason why \emph{precision} is not considered is that in this work we assume all the repackaged pairs in our benchmark are true positives. Therefore, there will be no false positives reported and hence the precision of state-of-the-art tools will be always 100\%.}.
If a pair in the RePack benchmark is flagged as a repackaged pair, we consider it as a True Positive (TP) result. Otherwise, if the pair is flagged as a non-repackaged pair, we consider it as a False Negative (FN) result.
Then, the recall of a given tool can be computed based on the following formula:

\begin{equation}\label{eq:recall}
\footnotesize
recall:=\frac{TP}{TP + FN}
\end{equation}

\begin{figure}[!th]
    \centering 
    \includegraphics[width=0.8\linewidth]{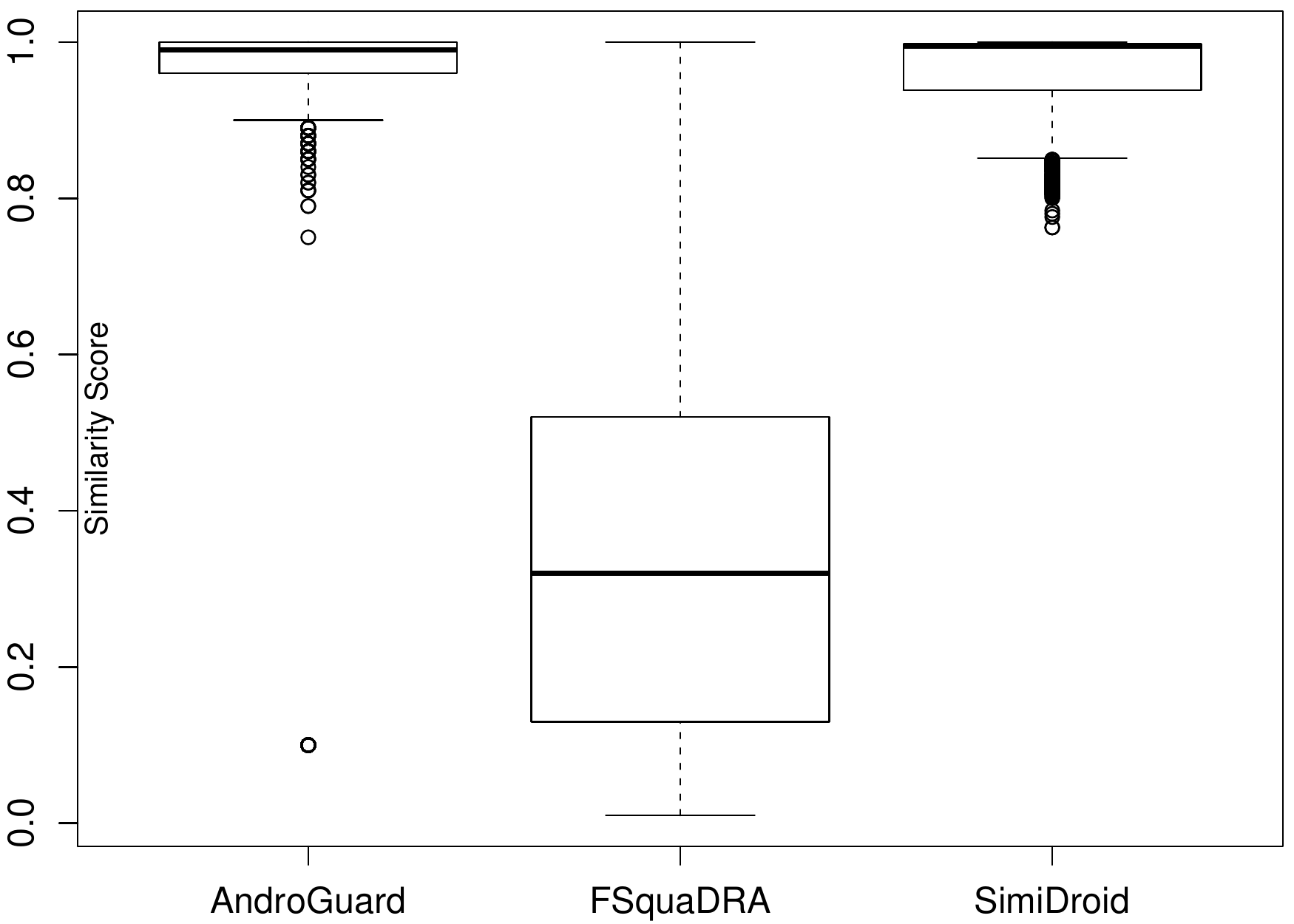}
    \caption{Distribution of Similarity Scores Given by AndroGuard, FSquaDRA and SimiDroid.}
    \label{fig:replication}
\end{figure}

Due to exceptions thrown by AndroGuard, FSquaDRA and SimiDroid, we eventually consider results for 8,078 pairs, among the pairs in \repo{}, where all the three tools have successfully finished their analyses.
Given the same threshold at \emph{80\%}, AndroGuard is in agreement with our collected dataset for 86\% pairs while FSquaDRA only agrees for 11\% pairs. In other words, while similarity scores by AndroGuard would allow identifying 86\% of pairs in our dataset (i.e., recall is 86\%), similarity by FSquaDRA is only aligned for 11\% of the \repo{} pairs (i.e., recall is 11\%).
Regarding the similarity results of SimiDroid, we only observe seven pairs (out of in total 11,255 successfully finished analyses) that have their similarity scores less than \emph{80\%}, resulting in almost 100\% recall.
Fig.~\ref{fig:fsquadra_thresholds} further plots the results of FSquaDRA by different thresholds.
The lower threshold considered, the higher the results achieved.
Nonetheless, even with lower thresholds, the results of FSquaDRA are still not comparable to that of AndroGuard and SimiDroid.
These results show that the similarity analysis we have performed for building possible repackaging pairs is highly in line with the analysis of AndroGuard and SimiDroid but not in line with the analysis of FSquaDRA.
The disagreement of FSquaDRA could be explained by one of the findings summarized by Li et al.~\cite{li2017understanding}, where the authors experimentally demonstrate that repackaging may also largely manipulate the resource files which would thus lead to poor results for resource-based similarity analysis tools.
Fig.~\ref{fig:replication} further shows the distribution of the similarity results of these three tools, where the median and mean values are 99\%, 85.8\% for AndroGuard, 30\%, 35\% for FSquaDRA, and  99.47\%, 96.78\% for SimiDroid, respectively.
Mann-Whitney-Wilcoxon (MWW) test demonstrates that the difference between the obtained scores of AndroGuard and FSquaDRA (and between that of FSquaDRA and SimiDroid) are significant.
Cliff's effective size estimation nevertheless suggests that the results of AndroGuard and FSquaDRA (and also that of SimiDroid and FSquaDRA) are largely and positively correlated, which has been also demonstrated by the authors of FSquaDRA.

Finally, we proceed ourselves to propose a straightforward scalable repackaged app detection approach based on our insights on analysing repackaging pairs in the \repo{} dataset. 
This approach uses classification technique based on symptoms (i.e., component capability declaration, app permission requests, mismatch between package and Launcher component names, package name diversity, favoured sensitive apis) of the repackaging process appearing in apps. 
Overall,  this straightforward approach achieves  81\%  F-Measure  for  distinguishing repackaged apps from non-repackaged ones.
Although  these  performance  scores  are  lower  than  many performances (up to 100\%) recorded in the literature, 
they are obtained, to the best of our knowledge, from the first readily reproducible  experiments  with  an  approach  that  is  not  based on pairwise similarity computation.
The performance of this technique, measured on \repo{}, can be considered as a reasonable baseline to improve within the community. 
For the sake of space, we provide the details of this approach as a supplementary document for interested readers to refer to.

\section{Priority Research Directions}
\label{sec:priority}
Our investigation into the problem of repackaged app detection has yielded an enumeration
of challenges that the research community must strive to address. We also propose in this 
section some priority research directions towards creating most efficient detection systems.

\textbf{1) Comprehending Repackaging Processes:}
As discussed previously, machine-learning approaches are appealing to ensure scalability and
practicality in repackaged app detection. Nevertheless, their design must be based on a solid
feature engineering process which should identify features that are truly representative
of the repackaging phenomenon. In this work, we have proposed preliminary investigations into
such features and focused on features which were easy and fast to extract.
More extensive analyses into a large ground truth of repackaging pairs can provide more insights
on novel and discriminative features. For example, we have not studied the impact of repackaging
on the density of call graphs (since edges are inserted/deleted by modifications of original apps).

A more extensive investigation into repackaging processes can further provide a taxonomy of repackaging similar to the types of code clones surveyed by Roy et al.~\cite{Roy:2009:CEC:1530898.1531101}. Such a taxonomy can help researchers more precisely inform readers on the target of their research.

\textbf{2) Graph Analysis:}
In static analysis, call/dependency graphs are known to be a reliable representation of app
structure. Since repackaging is mostly not invasive, a repackaged app may be formed by loosely-coupled modules~\cite{tian2016analysis}. An efficient analysis of call/ dependency graphs can thus be leveraged in a precise and practical means to detect repackaged apps, i.e., 
without relying on the availability of the original apps for comparison. 

\textbf{3) Dynamic Analysis:}
Dynamic analysis, although accurate, is expensive to implement by all market maintainers. 
Furthermore, because malware can now easily detect when they are run in a sandboxed environment, 
they may hide the behaviour implemented in the payload added via repackaging. Nevertheless, 
new avenues of research involved a form of crowdsourcing can be explored, where market maintainers collect runtime information from users' devices for posterior analyses of similar/divergent app behaviour. 

\textbf{4) Repackaging Deterrence:}
Finally, besides providing market maintainers with means to screen markets for repackaged apps, 
researchers should provide repackaged app detection in markets, developers should be provided with means 
for protecting their apps from repackagers. AppInk~\cite{zhou2013appink} and DIVILAR~\cite{zhou2014divilar}
are rare examples of research work attempting to propose approaches for repackaged deterrence in the
security community. We feel that the Software engineering community can heavily contribute in this
area as well with watermarking techniques.

\section{Threats to Validity}
\label{sec:validity}

Although we have tried our best to collect relevant papers as much as possible by following a well-defined methodology for conducting SLR, our results may have still missed some publications.
More specifically, the state-of-the-art repository search engines (e.g., the one provided by Springer) are not so accurate, usually resulting in many irrelevant papers and may also miss some relevant ones.
To further mitigate this threat, we have also conducted a backwards-snowballing based on already considered publications.

Another threat to validity of our study lies in the exhaustiveness of our dataset.
However, we have leveraged the AndroZoo largest available research dataset of Android apps to mitigate potential threats.
Nevertheless, we have empirically shown that our collected dataset, namely \repo{}, is quite diverse, e.g., 
containing both small-size and large-size apps and containing over 40 distinct types of repackaged malware families.

As a known challenge, so far, this is no straightforward means to pinpoint whether a given app is the original version of a given repackaging app pair. 
Hence, the original apps in our collected dataset may not be the final original ones as the creation time of DEX files can be manipulated.
Furthermore, as shown in a recent study, because of multi-generation repackaging, the identified original app of a given repackaging pair may also be a repackaged version of a previous app, making the identified original app even more wrongful.
Furthermore, the creation time of Android apps can be also manipulated, making the identified original app even more wrongful.
Nevertheless, it remains an interesting future work for our community to tackle.

Because of AV disagreements, the AV labels collected from VirusTotal may not be perfect, nor does the AVClass classifications.
However, in this work, we only use VirusTotal to get quick insights on our constructed dataset.
We thus encourage our fellow researchers to explore this direction to propose more promising approaches for pinpointing Android malware.

The fact that libraries are not excluded in this work could lead to inaccurate results as well.
The rationale behind this decision is that 
(1) Despite much efforts have been put on investigating Android libraries, we feel that the momentum of Android research has not yet produced a complete set of common libraries to support in-depth analysis of Android apps, including the whitelist leveraged in the extension of this work. 
(2) As argued by literature work, libraries could be favoured by attackers to inject malicious payloads so as to repackage Android apps with the ability to reuse the same exploitation to other apps that have leveraged the same popular library.
Removing libraries may also exclude the opportunity to pinpoint the malicious behaviours of repackaged malicious apps.
Nonetheless, we believe that the consideration of libraries could be crucial to repackaged app detection approaches. 
We, therefore, have provided to our community a research-based prototype tool for generating library-free versions of given Android apps, aiming at encouraging the community to innovate in-depth analyses for better understanding the facts between libraries and app clones, including malicious ones.

Finally, despite our benchmark is carefully built following a strict definition of repackaged app pairs: (1) over 80\% of code similarity and should be signed by different developers, our benchmark may lead to both under-approximate (obfuscation is not considered) and over-approximate (libraries are considered) results.
Indeed, under-approximate could be reported if obfuscation (especially method signatures are manipulated) is not considered while over-approximate could be yielded if common libraries are taken into account.
Additionally, the definition by itself may not be always true. 
For example, two apps could be independently implemented by two developers (i.e., different signatures) via cloning from the same app.
Because the changes made by the cloning process can be small, the two apps may still remain over 80\% of code similarity (mainly contributed by the app that the two apps cloned from).
Based on our definition, we could still flag this two apps as a repackaged app pair.
Nonetheless, although the repackaged app version is not directly modified from another app, we believe it could still be considered (to some extent) as a repackaged pair.
Moreover, the current construction process of the benchmark is based on the extracted properties at the method level, which might introduce biases to repackaged app detectors that are implemented based on other means rather than the static analysis of methods.
Nevertheless, it is non-trivial and probably time-intensive to manually verify the validity of all the app pairs in our benchmark.
Hence, we commit to continuously improve the validity of the benchmark and eventually provide an oracle for evaluating repackage detection tools.

\section{Related Work}
\label{sec:related_work}

Repackaging is an important issue in the Android ecosystem that must
be continuously dealt with by the research community. 
We argue 
that the research around repackaged app detection is blocked by state-of-the-art work
which record high performance rates in the literature while hindering
 comparative assessments. Related to our work are (1) studies that provide 
constructive discussions on the value of contributions in a research
domain, (2) general research on clone detection and (3) frameworks for
assessing repackaged detection approaches.

\paragraph*{Critical review of research}
Recently, Blackburn et al. present to our community a pragmatic guide to assessing empirical evaluations~\cite{blackburn2016truth}. 
They state that an unsound claim can misdirect an entire field. 
In this work, we attempt to follow their guidelines and thus to avoid potentially unsound claims.
Although we have not focused on finding fallacies in evaluation of state-of-the-art
work, our motivation is similar to that of Monperrus~\cite{Monperrus:2014:CRA:2568225.2568324} 
and his critical review of a state-of-the-art automated repair work. 
We have discussed the challenges that
researchers in this domain must keep in mind and further provide new data, approach and ideas for potential
research directions.

\paragraph*{Code clone \& software plagiarism detection}
App repackaged detection deals with similar concepts as in traditional 
code clone detection approaches~\cite{Roy07asurvey,Baker95,Baxter98,Jiang:2007,Liu06gplag:detection, Basit2009}, which
either work at a higher level where files are directly compared~\cite{Basit2009} or 
work at a lower level where fragments of code (or graphs/trees) are considered, their 
objectives were to measure the similarity of code fragments.
Even if the notion of clone fragment, be it a method, file or package, could be very useful for app similarity measurements, it is not sufficient in the context of Android, since Android apps have intensively leveraged framework and library code. 
In other words, two apps with similar code fragments are not necessarily similar. 

Closely related to our new proposed approach is Clonewise~\cite{Clonewise2013}, which, 
to the best of our knowledge, is the first to consider clone detection as a classification problem. 
Our approach, also in contrast with most state-of-the-art work, considers repackaging  
detection as a classification problem to enable a practical use in real-world settings.

Nevertheless, machine learning techniques, by allowing sifting through large sets of applications 
to detect malicious apps based on measures of similarity of features, 
have been extensively leveraged to conduct large-scale malware detection~\cite{Kolter:2006, Zhang:2007, Sahs:2012, McBoost}.
Unfortunately, through extensive evaluations, the community of ML-based malware detection has not yet 
shown that current malware detectors for Android are efficient in detecting malware in the wild. 
One among the candidate reasons to this situation is the fact that most malware are actually repackaged from benign apps, their ML-based features are probably similar to those extracted from benign apps, making them indistinguishable for ML-based malware detection.
Indeed, as pointed out by Meng et al.~\cite{meng2016semantic}, the current feature-based malware detection approaches are not enough because they cannot provide detailed information beyond their mere detection.
They thus propose an alternative approach that leverages semantic features (based on deterministic symbolic automaton (DSA)) to comprehensive Android malware and thereby to detect and classify them.
Therefore, we believe that the detection of repackaged Android apps contributes to also valuable ingredients for detecting malicious Android apps.

\paragraph*{Assessment of repackaged detection algorithms} Complementary to our work, Huang et al.~\cite{huang2013framework}
have early proposed a framework to comprehensively evaluate the obfuscation resilience 
of repackaging detection algorithms. They demonstrate the obfuscation problem for repackaged detection
algorithm by experimenting on Androguard. Following state-of-the-art work now regularly report on their
performance with this framework. With our work, we aim for the same momentum of using a common dataset
for evaluating approaches.

\section{Conclusion}
\label{sec:conclusion}
We proposed to review the challenges of repackaged app detection
in the Android ecosystem. We then performed a review of state-of-the-art work 
and highlighted the necessity to put new life into the research on repackaged app detection.
We contribute in this direction by building a comprehensive dataset of repackaging pairs, aiming at supporting replications of existing approaches and implications of new research directions.

\section*{Acknowledgment}
This work was supported by the Fonds National de la Recherche (FNR), Luxembourg, under the project CHARACTERIZE C17/IS/11693861 and Recommend C15/IS/10449467.
The authors would like to thank the anonymous reviewers who have provided insightful and constructive comments that have led to important improvements in several parts of the manuscript.
The authors also appreciate the help received from Timoth\'ee Riom who have helped verifying the literature review results.

\bibliographystyle{unsrt}
\bibliography{sa3} 

\end{document}